# Characteristics of *Kepler* planetary candidates based on the first data set: the majority are found to be Neptune-size and smaller

William J. Borucki[0,1], David G. Koch[1], Gibor Basri[2], Natalie Batalha[3], Alan Boss[4], Timothy M. Brown[5], Douglas Caldwell[6], Jørgen Christensen-Dalsgaard[7], William D. Cochran[8], Edna DeVore[6], Edward W. Dunham[9], Andrea K. Dupree[10], Thomas N. Gautier III[11], John C. Geary[10], Ronald Gilliland[12], Alan Gould[13], Steve B. Howell[14], Jon M. Jenkins[6], Hans Kjeldsen[7], David W. Latham[10], Jack J. Lissauer[1], Geoffrey W. Marcy[2], David G. Monet[15], Dimitar Sasselov[10], Jill Tarter[6], David Charbonneau[10], Laurance Doyle[6], Eric B. Ford[16], Jonathan Fortney[17], Matthew J. Holman[10], Sara Seager[18], Jason H. Steffen[19], William F. Welsh[20], Christopher Allen[21], Stephen T. Bryson[1], Lars Buchhave[10], Hema Chandrasekaran[6], Jessie L. Christiansen[6], David Ciardi[22], Bruce D. Clarke[6], Jessie L. Dotson[1], Michael Endl[8], Debra Fischer[23], Francois Fressin[10], Michael Haas[1], Elliott Horch[24], Andrew Howard[2], Howard Isaacson[2], Jeffery Kolodziejczak[25], Jie Li[6], Phillip MacQueen[8], Søren Meibom[10,] Andrej Prsa[26], Elisa V. Quintana[6], Jason Rowe[1], William Sherry[14], Peter Tenenbaum[6], Guillermo Torres[10], Joseph D. Twicken[6], Jeffrey Van Cleve[6], Lucianne Walkowicz[2], and Hayley Wu[6]

[1]NASA Ames Research Center, Moffett Field, CA 94035, USA
[2]University of California, Berkeley, CA, 94720, USA
[3]San Jose State University, San Jose, CA, 95192, USA
[4]Carnegie Institute of Washington, Washington, DC 20015 USA
[5]Las Cumbres Observatory Global Telescope, Goleta, CA 93117, USA
[6]SETI Institute, Mountain View, CA, 94043, USA
[7]Aarhus University, Aarhus, Denmark
[8]McDonald Observatory, University of Texas at Austin, Austin, TX, 78712, USA
[9]Lowell Observatory, Flagstaff, AZ, 86001, USA
[10]Harvard-Smithsonian Center for Astrophysics, Cambridge, MA, 02138, USA
[11]Jet Propulsion Laboratory, Calif. Institute of Technology, Pasadena, CA, 91109, USA
[12]Space Telescope Science Institute, Baltimore, MD, 21218, USA
[13] Lawrence Hall of Science, Berkeley, CA USA
[14]NOAO, Tucson, AZ 85719 USA
[15]United States Naval Observatory, Flagstaff, AZ, 86001, USA
[16]Univ. of Florida, Gainesville, FL, 32611 USA
[17]Univ. of Calif., Santa Cruz, CA 95064 USA
[18]MIT, Cambridge, MA 02139 USA
[19]Fermilab, Batavia, IL 60510 USA
[20]San Diego State Univ., San Diego, CA 92182 USA
[21]Orbital Sciences Corp., Mountain View, CA 94043 USA
[22]Exoplanet Science Institute/Caltech, Pasadena, CA 91125 USA
[23]Yale University, New Haven, CT 06520 USA
[24]Southern Connecticut State University, New Haven, CT 06515 USA
[25]MSFC, Huntsville, AL 35805 USA
[26]Villanova University, Villanova, PA 19085 USA

[0]Correspondence should be addressed to: William Borucki, William.J.Borucki@nasa.gov

**Abstract.** In the spring of 2009, the *Kepler* Mission commenced high-precision photometry on nearly 156,000 stars to determine the frequency and characteristics of small exoplanets, conduct a guest observer program, and obtain asteroseismic data on a wide variety of stars. On 15 June 2010 the *Kepler* Mission released data from the first quarter of observations. At the time of this publication, 706 stars from this first data set have exoplanet candidates with sizes from as small as that of the Earth to larger than that of Jupiter. Here we give the identity and characteristics of 306 released stars with planetary candidates. Data for the remaining 400 stars with planetary candidates will be released in February 2011. Over half the candidates on the released list have radii less than half that of Jupiter. The released stars include five possible multi-planet systems. One of these has two Neptune-size (2.3 and 2.5 Earth-radius) candidates with near-resonant periods.

## 1. Introduction

*Kepler* is a Discovery-class mission designed to determine the frequency of Earth-size planets in and near the habitable zone (HZ) of solar-type stars. The instrument consists of a 0.95 m aperture telescope/photometer designed to obtain high-precision photometric measurements of > 100,000 stars to search for patterns of transits. The focal plane of the Schmidt-type telescope contains 42 CCDs with a total of 95 megapixels that cover 115 square degrees of sky. *Kepler* was launched into an Earth-trailing heliocentric orbit on 6 March 2009, finished its commissioning on 12 May 2009, and is now in science operations mode. Further details of the *Kepler* Mission and instrument can be found in Koch *et al*. (2010b), Jenkins *et al*. (2010c), and Caldwell *et al*. (2010).

During the commissioning period, photometric measurements were obtained at a 30-minute cadence for 53,000 stars for 9.7 days. During the first 33.5 days of science-mode operation, 156,097 stars were similarly observed. Five new exoplanets with sizes between 0.37 and1.6 Jupiter radii and orbital periods from 3.2 to 4.9 days were confirmed by radial velocity observations (Borucki *et al*. 2010, Koch *et al*. 2010a, Dunham *et al*. 2010, Jenkins *et al*. 2010a, and Latham *et al*. 2010). Several hundred candidates were recognized, but there was not sufficient time to confirm more prior to the setting of the star field as seen by ground-based observers in late 2009. At the one-year anniversary of the receipt of the first set of data from the beginning of science operations, the data for 156,097 stars covering these two periods are now available to the public, apart for two exceptions: 400 stars held back to allow completion of one season of observations by the *Kepler* team, and 2778 stars held back for the Guest Observers and Asteroseismic Science Consortium (KASC). These data will be released on 1 February 2011, and in November 2010 when the proprietary period is complete, respectively. A total of 152,919 stars are now available at several levels of processing at the Multi-Mission Archive at the Space Telescope Science Institute (MAST[1]) for analysis by the community.

Because of great improvements to the data-processing pipeline, many more candidates are readily visible than in the data used for the papers published earlier this year. Over 850 stars with transiting exoplanet signatures have been identified. Of those, approximately 150 have been identified as likely false positives and, consequently, removed from consideration as viable exoplanet candidates. As false positives are confirmed, they will be archived at MAST. Four hundred of the 706 target stars with exoplanet candidates have been held back as the *Kepler* team conducts follow-up observations during the 2010 ground-based observing season. The discussion in this paper covers the remaining 306 stars that the *Kepler* team does not plan to give high priority for follow-up confirmation. These stars generally are associated with faint stars, and were not observed for the first 9.7-day time interval. Thus only 33.5 days of data are available for most candidates discussed herein. The characteristics of these candidates are presented and an appendix identifying these objects and providing their characteristics is attached. A separate paper that identifies false positive events found in the released data will be submitted. In the interim, see the list at the MAST. False positive events are patterns of dimming that appear to be the result of planetary transits, but are actually caused by other astrophysical processes or by instrumental fluctuations in the brightness values that mimic planetary transits. The identification of the false positives should help the community to avoid wasting observation resources.

[1] http://archive.stsci.edu/kepler/kepler_fov/search.php

The algorithm that searches for patterns of planetary transits also finds stars with multiple planet candidates. Several examples are shown in the section 4. A separate paper has been submitted that presents an analysis of these candidates (Steffen *et al*. 2010).

Data and search techniques capable of finding planetary transits are also very sensitive to eclipsing binary (EB) stars, and indeed the number of EBs discovered with *Kepler* vastly exceeds the number of planetary candidates. With more study, some of the current planetary candidates might also be shown to be EBs. Prsa *et al*. (2010) present a list of EBs with their basic system parameters that have been detected in these early data.

## 2. Description of the Data

The results discussed in this paper are based on the first two data segments taken at the start of the Mission. The first data segment is a 9.7-day period (labeled Q0) starting on 2 May 2009 UT that occurred during the commissioning phase. The second is a 33.5-day segment (labeled Q1) that started at the beginning of science operations on 13 May 13 2009 UT and finished on 15 June 2009 UT. Both periods occurred at the initial orientation of the spacecraft. During Q0, all the sufficiently un-crowded stars in the field of view (FOV) brighter than 13.6 and fainter than 5 in the *Kepler* passband (Kp) were included. A total of 52,496 stars including both dwarfs and giants, were observed. The *Kepler* Kp band pass cover both the V and R photometric pass bands.

The Q1 observations used *Kepler*'s normal list of 156,097 exoplanet target stars. These stars are primarily main sequence dwarfs chosen from the *Kepler* Input Catalog[27] (KIC). Stars were chosen to maximize the number of stars that were both bright and small enough to show detectable transit signals for small planets in and near the HZ (Batalha *et al*. 2010b). Most stars were in the magnitude range $9 < Kp < 16$.

Data for all stars are recorded at a cadence of one per 29.4 minutes (hereafter, long cadence, or LC). Data for a subset of 512 stars are also recorded at a cadence of one per 58.5seconds (hereafter, short cadence or SC), sufficient to conduct asteroseismic observations needed for measurements of the stars' size, mass, and age. The results presented here are based only on LC data. For a full discussion of the LC data and their reduction, see Jenkins *et al*. (2010b, 2010c). See Gilliland *et al*. (2010) for a discussion of the SC data.

### *2.1 Noise Sources in the Data*

The *Kepler* photometric data contain a wide variety of both random and systematic noise sources. Random noise sources such as shot noise from the photon flux and read noise have (white) Gaussian distributions. Stellar variability introduces red (correlated) noise. For many stars, stellar variability is the largest noise source. There are also many types of instrument-induced noise: pattern noise from the clock drivers for the "fine-guidance" sensors, start-of-line ringing, overshoot/undershoot due to the finite bandwidth of the detector amplifiers, and signals that move through the output produced by some of the amplifiers that oscillate. The latter noise patterns (which are typically smaller than one least-significant-bit in the digital-to-analog converter for a single read operation) are greatly affected by slight temperature changes, making their removal difficult. Noise due to pointing drift, focus changes, differential velocity aberration, CCD defects, cosmic ray events, reaction wheel heater cycles, breaks in the flux time series due to desaturation of the reaction wheels, spacecraft upsets, monthly rolls to downlink the data, and quarterly rolls to re-orient the spacecraft to keep the solar panels pointed at the Sun are also present. These sources and others are treated in Jenkins *et al*. (2010b) and Caldwell *et al*. (2010). Work is underway to improve the mitigation and flagging of the affected data. Additional noise sources are seen in the short cadence data (Gilliland *et al*. 2010). In particular, a frequency analysis of these data often shows spurious regularly spaced peaks at 48.9388 $day^{-1}$ and its harmonics. Additionally, there appears to be a noise source that causes additive offsets in the time domain inversely proportional to stellar brightness.

[27] http://archive.stsci.edu/*Kepler*/*Kepler*_fov/search.php

Because of the complexity of the various small effects that are important to the quality of the *Kepler* data, prospective users of *Kepler* data are strongly urged to study the data release notes (hosted at the MAST) for the data sets they intend to use. Note that the *Kepler* data analysis pipeline was designed to perform differential photometry to detect planetary transits so other uses of the data products require caution.

*2.2 Distinguishing Planetary Candidates from False Positive Events*

Stars that show a pattern consistent with those from a planet transiting its host star are labeled “planetary candidates.” Those that have failed some consistency test are labeled “false positives”. Thus the search for planets starts with a search of the time series of each star for a pattern that exceeds a detection threshold commensurate with a non-random event. After passing all consistency tests described below, and only after a review of all the evidence by the entire *Kepler* Science Team, does the candidate become a validated exoplanet. It is then submitted to a peer-reviewed journal for publication.

There are two general types of processes associated with false positive events in the *Kepler* data that must be evaluated and eliminated before a candidate planet can be considered a valid discovery: 1) statistical fluctuations or systematic variations in the time series, and 2) astrophysical phenomena that produce similar signals. A sufficiently high threshold has been used that statistical fluctuations should not contribute to the candidates proposed here. Similarly, systematic variations in the data have been interpreted in a conservative manner and only rarely should result in false positives. However, astrophysical phenomena that produce transit-like signals will be much more common.

*2.3 Search for False Positives in the Output of the Data Pipeline*

The *Kepler* data-processing pipeline reduces the photometric data for each star and then searches each time series for “threshold crossing events” (TCEs), a pattern of events that exceed a 7.1-$\sigma$ threshold (Batalha *et al*. 2010b, Jenkins *et al.* 2010a) and might be caused by planetary transits. After identification as a TCE, the following validation process is normally executed by the science team (Batalha et al. 2010a):

The photometric data are processed to remove trends due to instrumental effects and/or stellar variability. The data are folded according to a putative orbital period and an analytic model is fit to estimate the depth, duration, and shape of the possible transit. The duration, depth, and shape of the light curve must be appropriate for an orbiting companion. The transit depth and duration must be constant. The duration must also be consistent with the orbital period, estimated stellar mass and *Kepler*’s laws, assuming a small eccentricity.

Using these estimates and information about the star from the KIC, tests are performed to search for a difference in even- and odd-numbered event depths. If a significant difference exists, this would suggest that a comparable-brightness EB has been found for which the true period is twice that determined due to the presence of primary and secondary eclipses. Similarly, a search is conducted for evidence of a secondary eclipse or a possible planetary occultation roughly half-way between the potential transits. If a secondary eclipse is seen, then this could indicate that the system is an EB with the period assumed. However, the possibility of a self-luminous planet (as with HAT-P-7; Borucki *et al.* 2009) must be considered before dismissing a candidate as a false positive.

The shift in the centroid position of the target star measured in and out of the transits must be consistent with that predicted from the fluxes and locations of the target and nearby stars.

After passing these tests, the candidate is elevated to “Kepler Object of Interest” (KOI) status and is forwarded to the Follow-up Observation Program (FOP) for various types of observations and additional analysis. These observations include:

1. High resolution imaging with adaptive optics or speckle interferometry to evaluate the contribution of other stars to the photometric signal and to evaluate the shift of the photocenter when a transit occurs.

Medium-precision radial velocity (RV) measurements are made to rule out stellar or brown dwarf mass companions and to better characterize the host star.

A stellar –blend model (Torres *et al*. 2004) is used to check that the photometry is consistent with a planet orbiting a star rather than the signature of a multi-star system.

High-precision RV measurements may be made, as appropriate, to verify the phase and period of the most promising candidates and ultimately to determine the mass and eccentricity of the companion and to identify other non-transiting planets. For low-mass planets where the RV precision is not sufficiently high to detect the stellar radial velocity variations, RV observations are conducted to produce an upper limit for the planet mass and assure that there is no other body that could cause confusion.

When the observations indicate that the Rossiter-McLaughlin effect (Winn 2007) will be large enough to be measured in the confirmation process, such measurements may be scheduled, typically at the Keck Observatory.

When the data indicate the possibility of transit timing variations large enough to assist in the confirmation process, the multiple-planet and transit-timing working groups perform additional analysis of the light curve and possible dynamical explanations (Steffen *et al*. 2010).

This paper discusses the characteristics for the 306 candidates in the released list. Note that these candidates have not been fully vetted by the processes described above and that the false alarm rate for the candidates could be near 50% (Gautier *et al*. 2010).

## 3. Results

For the released candidates, the KOI number, the KIC number, the stellar magnitude, effective temperature, and surface gravity of the star taken from the KIC are listed in the Appendix. Also listed are the orbital period, epoch, and an estimate of the size of the candidate. More information on the characteristics of each star can be obtained from the KIC. Several of the target stars show more than one series of planetary transit-like events and therefore probably have more than one planetary candidate orbiting the star. These candidate multi-planet systems are of particular interest because it is unlikely that all of the candidate planets associated with a multiple-transiting candidate star can be false positives. The candidate multiple-planet systems (i.e., KOI 152, 191, 209, 877, and 896) are discussed in a later section.

### *3.1 Naming Convention*

It is expected that many of the candidates listed in the attached Appendix will be followed up by members of the science community and that many will be confirmed as planets. To avoid confusion in naming them, it is suggested that the community refer to *Kepler* stars as KIC NNNNNNN (with a space between the "KIC" and the number), where the integer refers to the ID in the *Kepler* Input Catalog archived at MAST. For planet identifications, a letter designating the first, second, etc. confirmed planet as "b", "c", etc. should follow the KIC ID number. At regular intervals, the literature will be combed for planets found in the *Kepler* star field, sequential numbers assigned, the IAU-approved prefix ("*Kepler*") added, and the information on the planet with its reference will be placed in the *Kepler* Results Catalog. Preliminary versions of this catalog will be available at the MAST and revised on a yearly basis.

### *3.2 Statistical Properties of Planet Candidates*

We have conducted some statistical analyses of the 306 released candidates to investigate the general trends and initial indications of the characteristics of the detected planetary candidates. The readers are cautioned that the sample that we are studying contains many poorly quantified biases. In particular, some of the released candidates could be false positives. Further, those candidates orbiting stars brighter than 13.9 magnitude and the small-size candidates (i.e., those with radii less than 1.25 $R_\oplus$), are not among the released stars. Nevertheless, the large number of candidates provide interesting, albeit tentative, associations with stellar characteristics. Comparisons are limited to orbital periods of 30 days; complete for two transits only to 17 days.

In the figures below, the distributions of various parameters are plotted and compared with values in the literature and those derived from the Extrasolar Planet Encyclopedia[2] (updated as of 14 May 2010).

The results discussed here for the 306 stars are largely based on the observations of 83,872 stars fainter than 14th magnitude and with effective temperatures ($T_{eff}$) greater than 3850K. Stellar parameters are based on KIC data. The function of the KIC was to provide a target sample with a low fraction of evolved stars that would be unsuitable for transit work, and to provide a first estimate of stellar parameters that is intended to be refined spectroscopically for KOI at a later time. Spectroscopic observations have not been made for the released stars, so it is important to recognize that some of the characteristics listed for the stars are uncertain, especially surface gravity (i.e., log *g*) and metallicity ([M/H]). The errors in the star diameters can reach 25%, with proportional changes to the estimated diameter of the candidates. For some planet candidates only one transit has been observed, so their orbital periods were estimated based on the transit duration and the assumptions of zero eccentricity and a central transit. Such orbital periods are very uncertain.

In figure 1, the distributions of magnitude and effective temperature are given for reference. In later figures, the association of the candidates with these properties is examined.

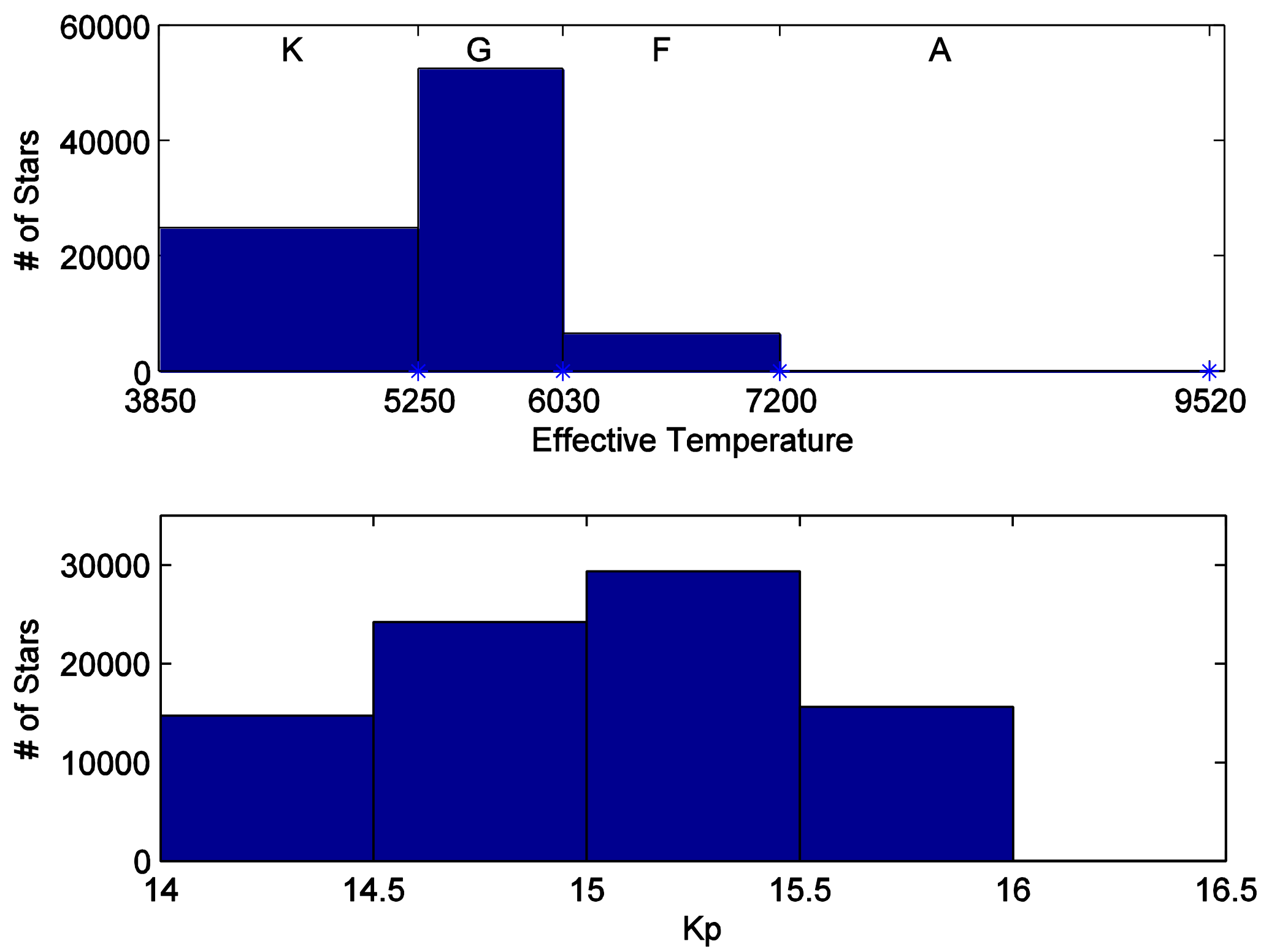

**Figure 1. Distributions of effective temperature and magnitude for the stars considered in this study. Letters K, G, F, and A refer to stellar spectral type.**

[2] Extrasolar Planet Encyclopedia; http://exoplanet.eu/

It is clear from figure 1 that most of the stars monitored by *Kepler* are G and K spectral types. This is because these types are the most frequent for a magnitude-limited survey of dwarfs and because the selection of target stars was purposefully skewed to enhance the detectability of Earth-size planets by choosing those with an effective temperature and magnitude that maximized the transit signal-to-noise ratio (SNR) (Batalha *et al*. 2010b). Thus, the decrease in the number of monitored stars for magnitudes greater than 15.5 is due to the selection of only those stars in the FOV that are likely to be small enough to show planets. In particular, A, F, and G stars were selected at magnitudes where they are sufficiently bright for their low shot noise to overcome the lower SNR for a given planet size due to their large stellar radii. After all available bright dwarf stars are chosen for the target list, many target slots remain, but only dimmer stars are available (Batalha *et al*. 2010b). From the dimmest stars, the smallest stars are given preference. In the following figures, when appropriate, the results will be based on the ratio of the number of candidates to the number of stars in each category.

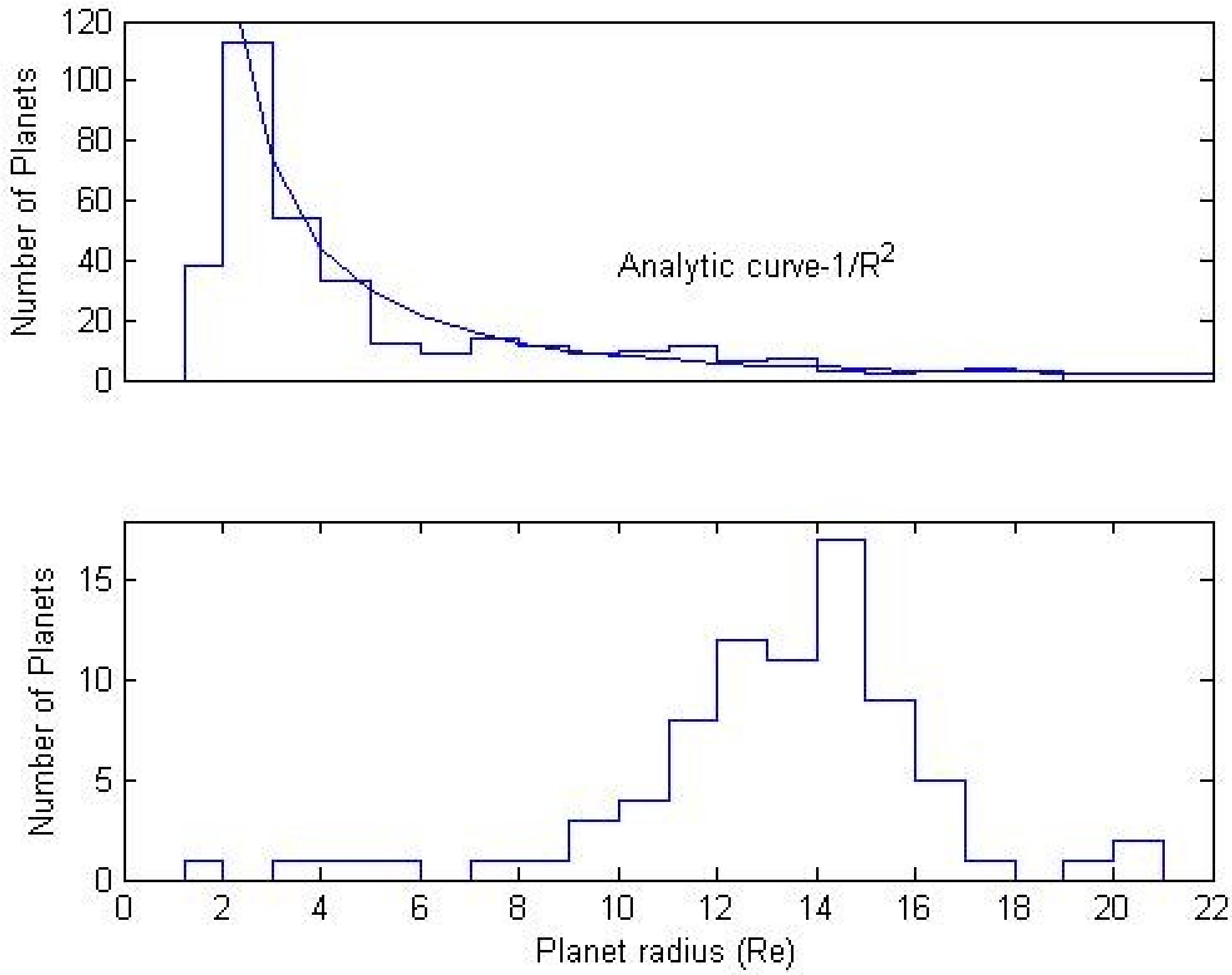

**Figure 2. Size distribution of *Kepler* candidates vs. planet radius (R) (upper panel). Size distribution of transiting planets listed in the Extrasolar Planet Encyclopedia (lower panel).**

A comparison of the distributions shown in figure 2 indicates that the majority of the candidates discovered by *Kepler* are Neptune-size (i.e., 3.8 $R_\oplus$) and smaller while the planets in the Extrasolar Planet Encyclopedia (EPE) are typically Jupiter-size (i.e., 11.2 $R_\oplus$) and larger. This difference is understandable because of the difficulty of detecting small planets when observing through the Earth's atmosphere and because of the inflation of highly irradiated planets that occurs for inner-orbit planets.

The *Kepler* results shown in figure 2 imply that small candidate planets with periods less than 30 days are much more common than large candidate planets with periods less than 30 days and that the ground-based discoveries are sampling the extended upper tail of the size distribution (Gaudi 2005). Note that for a substantial range of planet sizes, a $R^{-2}$ curve fits the *Kepler* data well. Because it is much easier to detect larger candidates than smaller ones, this result implies that the frequency of planets decreases with the area of the planet, assuming that the false positive rate and other biases are independent of planet size for planets larger than 2 Earth radii.

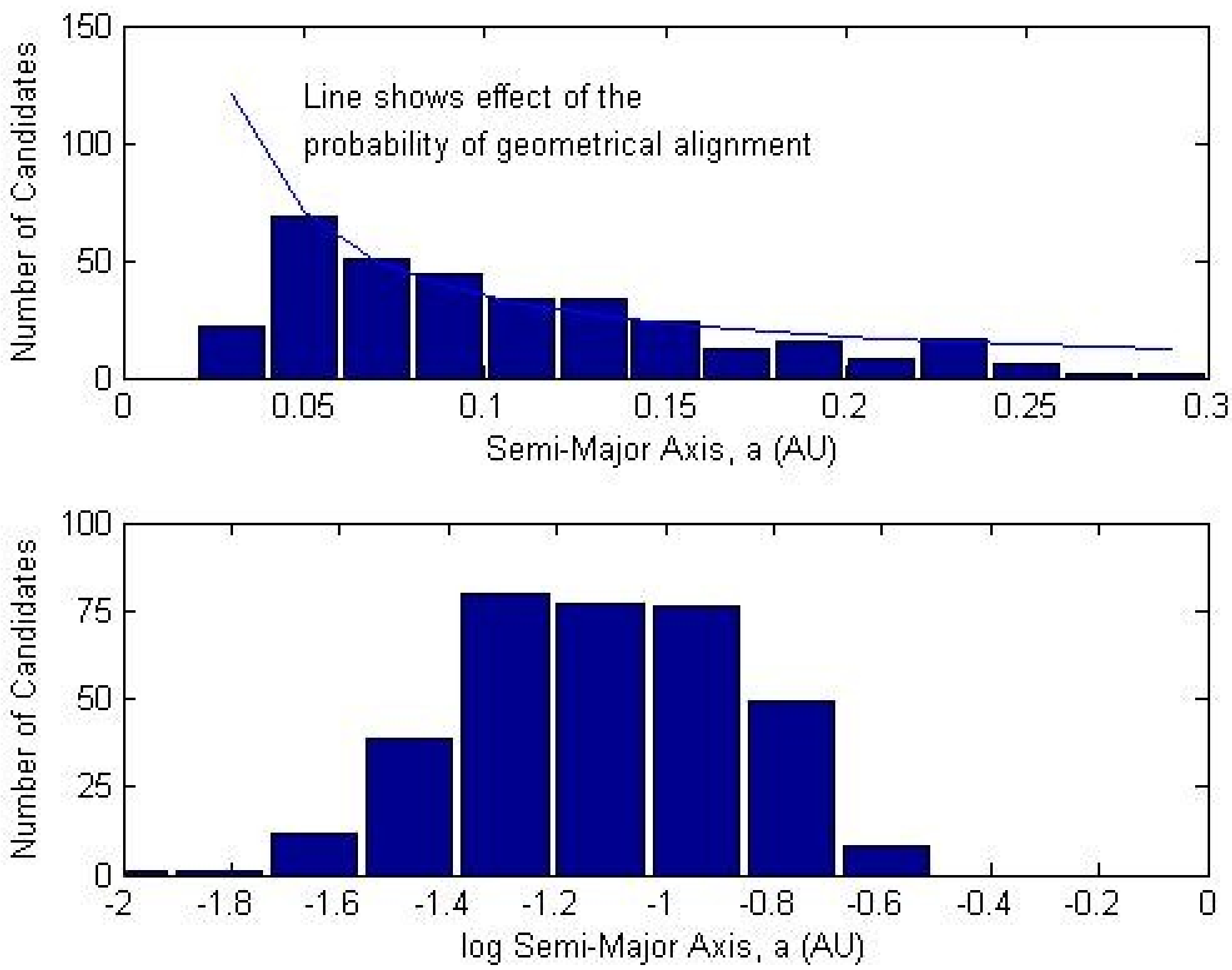

**Figure 3. Upper panel: The number of candidates vs. linear intervals in the semi-major axis . Lower panel: The number of candidates vs. logarithmic intervals of the semi-major axis.**

In figure 3, the dependence of the number of candidates on the semi-major axis is examined. In the upper panel an analytic curve has been fitted to show the expected reduction in the integrated number in each interval due to the decreasing geometrical probability that orbits are correctly aligned with the line-of-sight. It has been fit to the value at 0.05 AU and assumes that the number of candidates per linear interval in the semi-major axis is constant. The fit is consistent with the observations for semi-major axes < 0.25 AU, but it predicts values above those observed for larger values of the semi-major axis. It is possible that for the large values of semi-major axes, the short duration of the data string is causing some candidates to be missed. Since the requirement for a minimum of three transits was not imposed, periods out to 43 days should be present, with a gradual loss of candidates for periods exceeding 17 days (or a ~0.13 AU).

The lower panel of figure 3 presents the number of candidates in equal intervals of the logarithm of the semi-major axis. The values in the histogram would be level if the number of candidates per logarithmic interval were constant and the effect of the decreasing number expected from the dependence on the geometrical alignment probability was not present. The observations indicate that the hypothesis of equal numbers in equal logarithmic intervals is not supported. Appropriate corrections for the reduction in the number per interval due to dependence of the geometric probability did not change the situation. Thus the distribution of planet candidates does not appear to be consistent with the hypothesis of equal candidates in equal intervals of log semi-major axis.

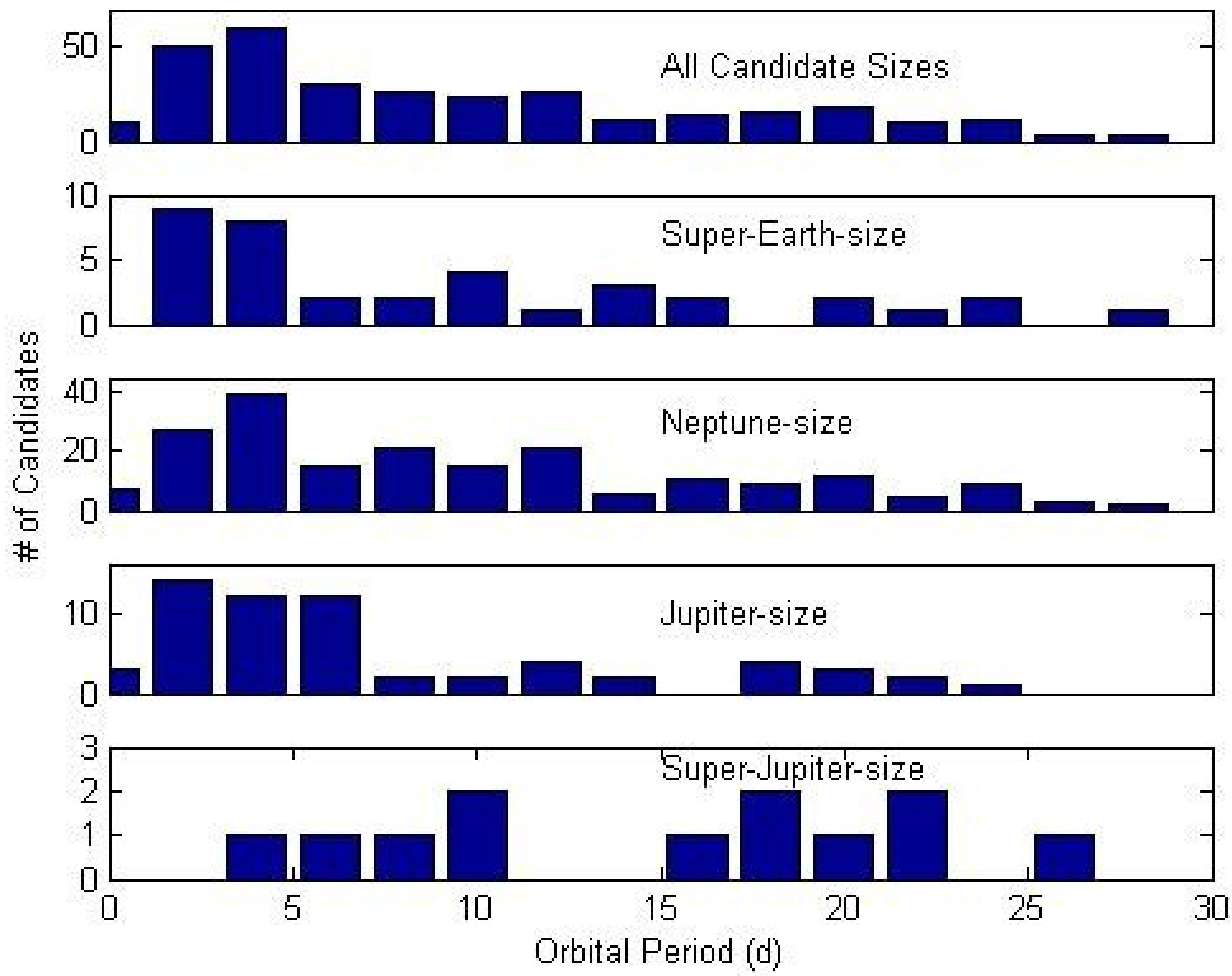

**Figure 4. Number of candidates vs orbital period for several choices of candidate size. Upper panel to lower panel: all candidate sizes, super-Earth-size candidates, Neptune-size candidates, Jupiter-size, and super-Jupiter-size candidates.**

In figure 4, each panel presents the orbital distribution for candidate sizes ranging from super-Earth to super-Jupiter. "super-Earth-size" candidates are those with sizes from 1.25 $R_\oplus$ to 2.0 $R_\oplus$. These are expected to be rocky type planets without a hydrogen-helium atmosphere. "Neptune-size" candidates are those with sizes from 2.0 $R_\oplus$ to 6 $R_\oplus$, and are expected to be similar to Neptune and the ice giants in composition. Candidates with sizes between 6 and 15 $R_\oplus$ and between 15 and 22 $R_\oplus$ are labeled Jupiter-size and super-Jupiter-size candidates, respectively. The nature of the larger category of objects is unclear. No mass measurements are available. It is possible that they are small stars transiting large stars. It is also possible that these are ordinary jovian planets whose stars have incorrect KIC radii.

The middle three panels in figure 4 indicate that the number of candidates is decreasing with orbital period regardless of size and that there is a peak in concentration for orbital periods between 2 to 5 days.

There are several references in the literature to the pile-up of giant planet orbital periods near 3 days (Santos and Mayor 2003) and a "desert" for orbital periods in excess of 5 days. Figure 5 is a comparison of distributions of frequency with orbital period for the *Kepler* results with that derived from the planets listed in the EPE. In this instance, the much larger number of planets listed in the EPE under RV discoveries was used in the comparison. The very compact distribution of frequency with orbital periods near 3 days seen in the EPE results is also seen in the *Kepler* results. However, there is little sign of the "desert" that has been discussed in the literature with respect to the RV results. We note that the *Kepler* sample contains a much larger fraction of super-Earth-size candidates than does the EPE sample and has a much better phase coverage.

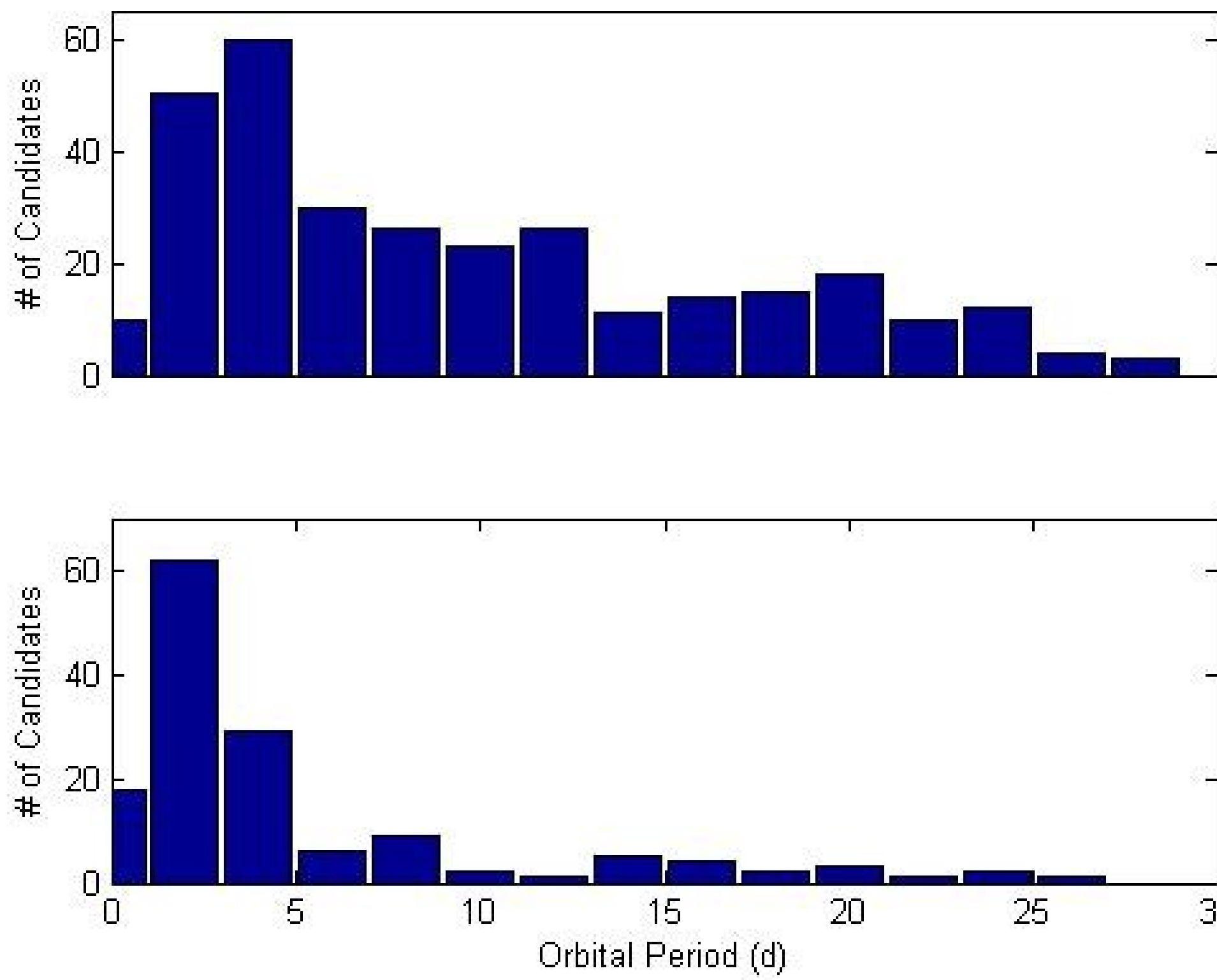

**Figure 5. Period distribution of candidate planets with planets (upper panel). Exoplanets listed in the EPE (updated 14 May 2010) determined from RV measurements (lower panel).**

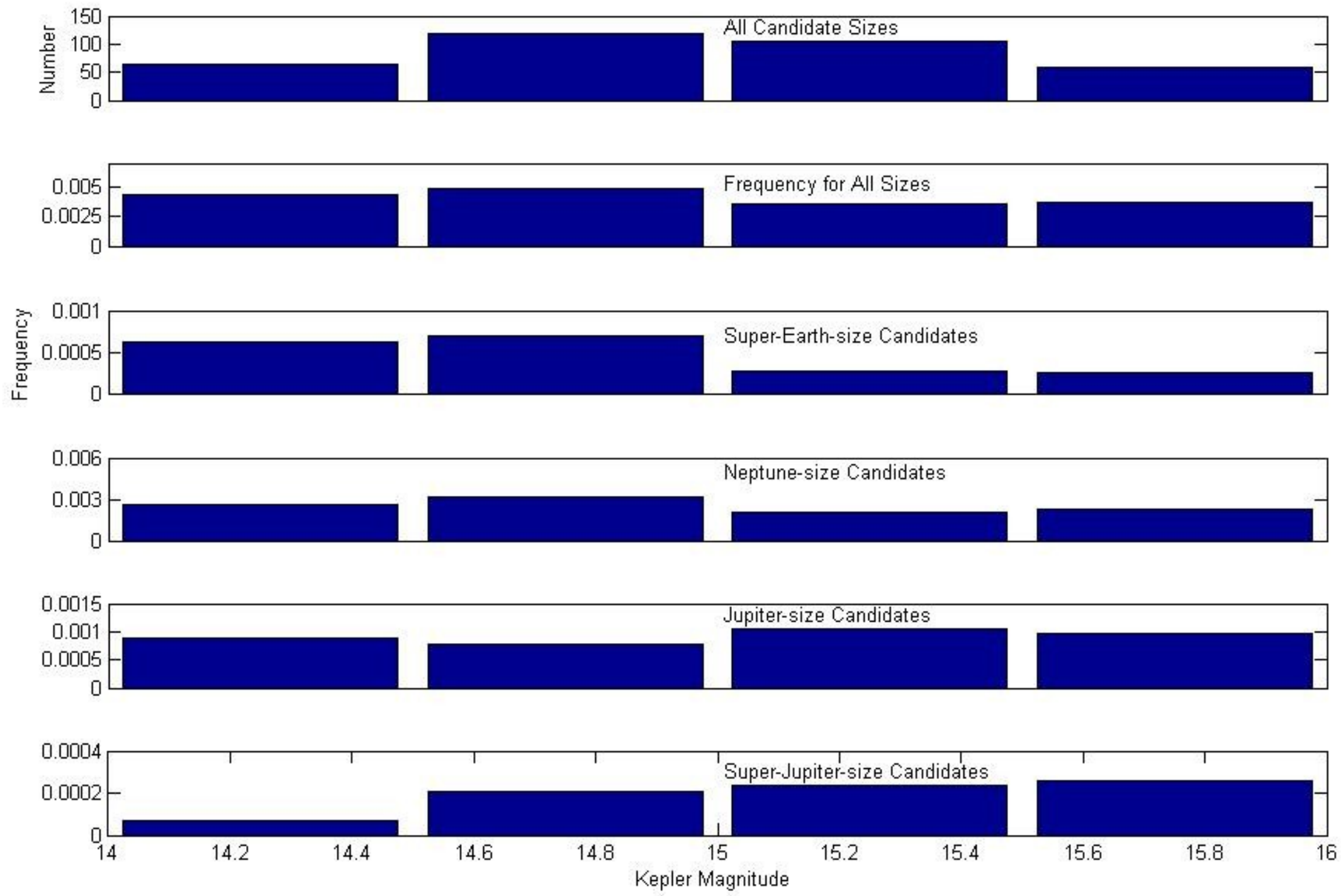

**Figure 6. Frequency (i.e., normalized fraction) of the target stars with candidates vs. magnitude. All magnitudes discussed here are *Kepler* magnitudes: Kp.**

In figure 6, the frequency of candidates in each magnitude bin has been calculated from the number of candidates in each bin divided by the total number of stars monitored in each bin. The numbers of stars brighter than 14.0 magnitude and fainter than 16.0 in the current list are so small that their counts are not shown.

The top panel shows that the number of candidates decreases for magnitudes larger than 15.0. This result is a combination of the bias caused by choosing brighter stars as targets and the decreased detectability of small candidates orbiting dim stars. The latter effect is especially noticeable for the frequency of the smallest candidate size shown (third from top panel). The observed drop-off in frequency of super-Earth-size candidates for magnitudes larger than 15 suggests that the survey is no longer complete above that magnitude. An alternative interpretation is that such planets are less common around the lower-mass stars that represent a larger fraction of our faint stars. The nearly constant values for the frequencies of Neptune- and Jupiter-size candidates indicates that their frequencies are independent of the stellar magnitude, as should be expected from their very large SNR.

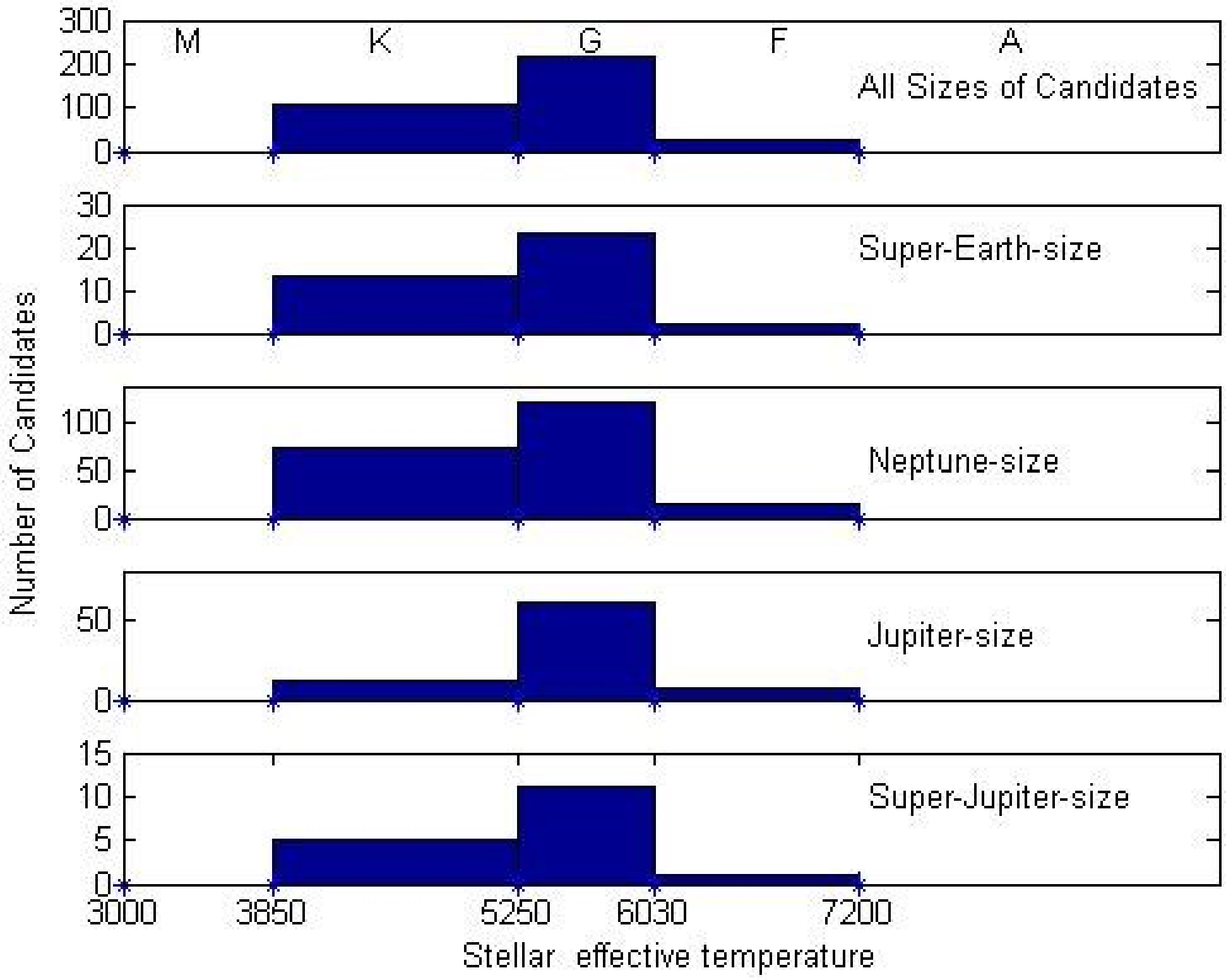

**Figure 7. Number of candidates for various candidate sizes vs. stellar effective temperature. The letters indicate spectral types of host stars.**

Figure 7 shows that the spectral types that produce most of the candidates are the G and K types. This result should be expected due to the large number of G- and K-type stars chosen as target stars. Note that the relatively large number of super-Earth candidates orbiting K stars is likely the result of small planets being easier to detect around small stars than around large stars.

In figure 8, the bias associated with the number of target stars monitored as a function of spectral type is removed by computing the frequencies of the candidates as a fraction of the number of stars monitored.

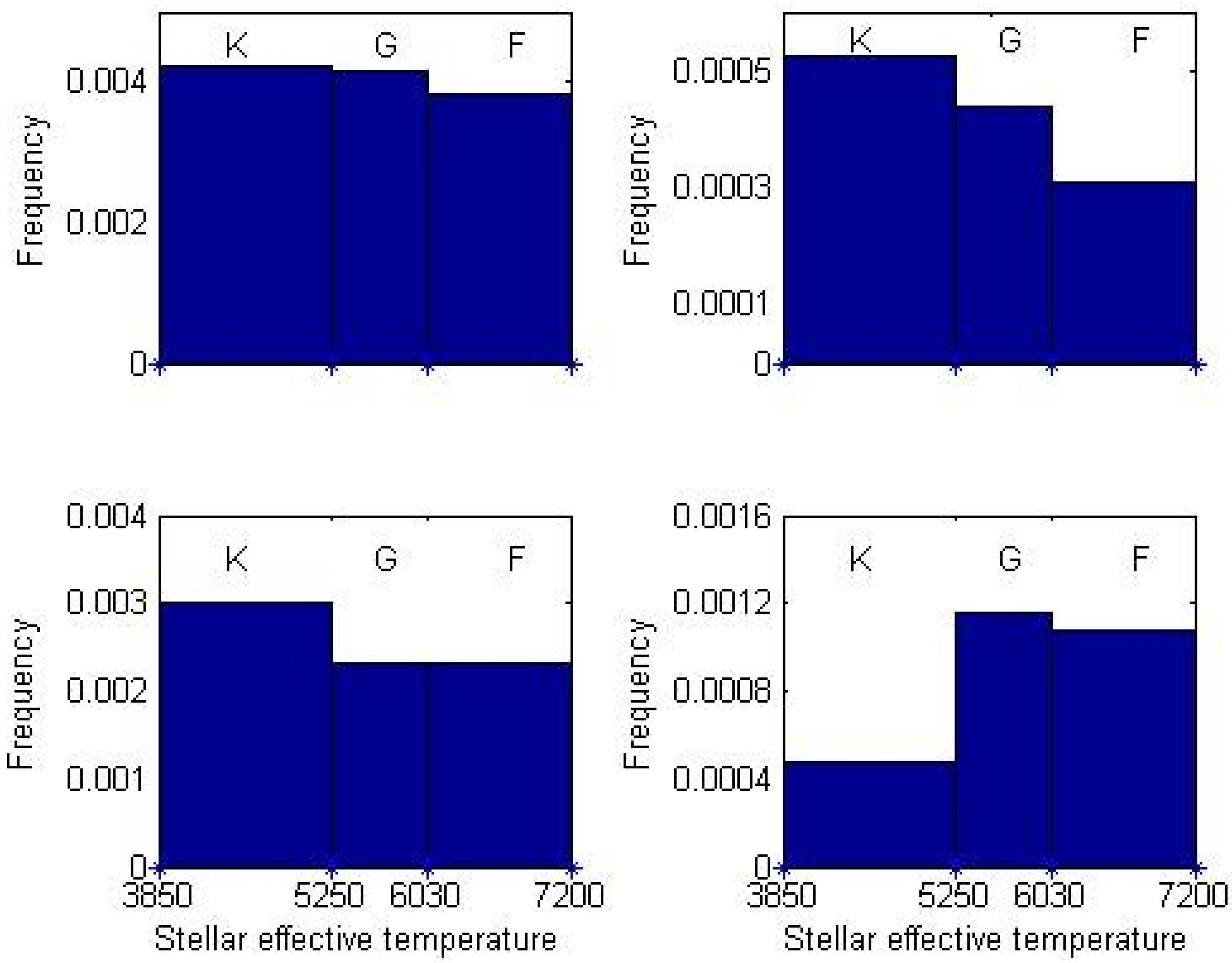

**Figure 8. Frequency of stars with candidates versus stellar effective temperature. Letters indicate the stellar spectral type. From upper left clockwise; All released candidates, super-Earth-size candidates, Neptune-size candidates, and Jupiter-size candidates.**

Note that the frequency for the total of all candidate sizes decreases only modestly with increasing stellar temperature. However, for super-Earth candidates, the decrease with temperature is quite marked, as might be expected when considering the substantially lower SNR due to the increase of stellar size with temperature. Thus it is unclear whether the decrease in occurrence frequency is real. The increase in the frequency for Jupiter-size candidates should not be biased the increasing stellar size because the signal level for such large candidates is many times the noise level associated with the instrument and shot noise. Thus, this increase is likely to be indicative of a real, positive association of giant candidates with stellar mass. Note that a study based on the observed correlation of the variation of total irradiance with the CaII index of chromospheric activity by Aigrain *et al*. (2004) predicted that the K dwarfs would be the most likely per star to show planets, followed by the Gs and then the Fs. It concluded that the K dwarfs would be more variable than the G dwarfs, but that their smaller diameters would more than compensate for the higher level of noise at the periods of interest to detecting transits.

A study of the dependence of the frequency of the planet candidates on the stellar metallicity was considered, but rejected because the metallicities in the KIC are not considered sufficiently reliable. In particular, the D51 filter used in the estimation of metallicity is sensitive to a combination of the effects of surface gravity and metallicity, especially within the temperature range from roughly late K to late F stars. However the information generated by this filter was used to develop the association with log g, thus making any estimate of metallicity highly uncertain.

An examination of figures 6 and 8 shows that the measured occurrence frequencies of candidate planets are somewhat dependent on the size of the candidates. Super-Earth-size candidates have an average frequency of about $5 \times 10^{-4}$ for brighter stars ($14.0 \leq Kp < 15.0$) while Neptune-size candidates have a lower frequency of $3x10^{-3}$ for such stars. Jupiter-size candidates have an even lower frequency of about $0.9x10^{-3}$, independent of stellar magnitude. Figure 8 indicates a positive correlation for Jupiter-size candidates with stellar effective temperature.

## 4. Examples of Candidate Multi-planet Systems

A number of target stars with multiple planet candidates orbiting a single star have been detected in the *Kepler* data. The light curves for five multi-candidate systems in the released data are shown in figures 9 through 13. Only a single transit-like event is seen for some planet candidates, as expected for planets with orbital periods exceeding the ~33 days of observations. For other candidate systems, several transits of multiple planet candidates have been observed.

In two cases, the ratio of putative orbital periods is near 2. For such a system there is a high (60%) conditional probability that both planets transit, provided that the inner planet transits and the system is planar. For systems with planet candidates having a large ratio of orbital periods (e.g., KOI 191), the probability that the outer planet will transit, given that the inner one does, is small. While an exhaustive study remains to be done, the implication is that many planetary systems have multiple planets or that nearly coplanar planetary systems might be common.

Any of these multiple planet candidate systems, as well as the single-planet candidate systems, could harbor additional planets that do not transit and therefore are not seen in these data. Such planets might be detectable via transit timing variations of the transiting planets after several years of *Kepler* photometry (Agol *et al*. 2005, Holman and Murray 2005). Based on the data presented here, we do not find any statistically significant transit timing variations for the five candidate multiple-planet systems or for the single planet candidates listed in the Appendix.

Table 1 lists the general characteristics of the five multi-candidate systems in the released data. It should be noted that in previous instances, multiple eclipsing binaries have been seen in the same photometric aperture and can appear to be multiple-planet systems. A thorough analysis of each system and a check for background binaries are required before any discovery should be claimed. A more extensive discussion of these candidates can be found in Steffen *et al.* (2010).

**Table 1. Properties of five multi-candidate systems.**

| ---------------Candidate Properties--------------- | | | | ------------ Stellar Properties--------------------------- | | | | |
|---|---|---|---|---|---|---|---|---|
| KOI # | Candidate size | Period (days) | Epoch[29] | KIC # | Teff[28] (K) | Kp | RA (2000) | Dec |
| 152.01 | 0.58 Rj | >27 | 91.747 | 8394721 | 6500 | 13.9 | 20 02 | 44 22 |
| 152.02 | 0.31 Rj | 27.41 | 66.630 | | | | 04.1 | 53.7 |
| 152.03 | 0.30 Rj | 13.48 | 69.622 | | | | | |
| 191.01 | 1.06 Rj | 15.36 | 65.385 | 5972334 | 5500 | 15.0 | 19 41 | 41 13 |
| 191.02 | 2.04 $R_\oplus$ | 2.42 | 65.50 | | | | 08.9 | 19.1 |
| 209.01 | 1.05Rj | >29 | 68.635 | 10723750 | 6100 | 14.2 | 19 15 | 48 02 |
| 209.02 | 0.68 Rj | 18.80 | 78.822 | | | | 10.3 | 24.8 |
| 877.01 | 2.53 $R_\oplus$ | 5.95 | 103.956 | 7287995 | 4500 | 15.0 | 19 34 | 42 49 |
| 877.02 | 2.34 $R_\oplus$ | 12.04 | 114.227 | | | | 32.9 | 29.9 |
| 896.01 | 0.38 Rj | 16.24 | 108.568 | 7825899 | 5000 | 15.3 | 19 32 | 43 34 |
| 896.02 | 0.28 Rj | 6.31 | 107.051 | | | | 14.9 | 52.9 |

[29] Epochs are BJD-2454900.

[28] The effective temperatures were derived from spectroscopic observations as described in Steffen *et al*. (2010).

### *4.1 Light Curves for Multi-candidate Systems*

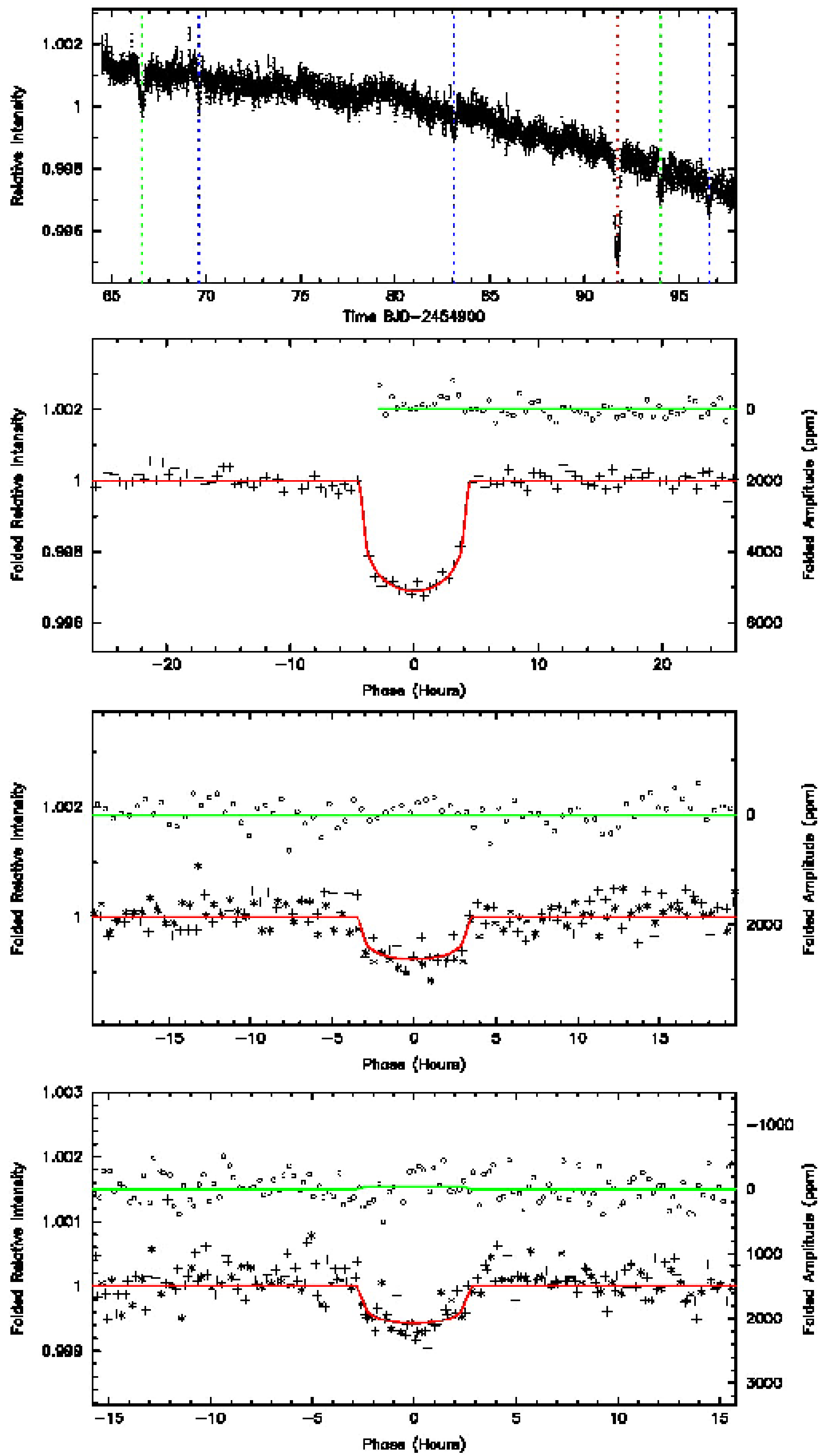
Relative Intensity
1.002
1
0.998
0.996
65
70
75
80
85
90
95
Time BJD−2454900
Folded Relative Intensity
Folded Amplitude (ppm)
Phase (Hours)
−20
−10
0
10
20
−15
−5
5
15
1.003
1.001
0.999
−1000
1000
2000
3000
4000
6000

**Figure 9. The three candidate planets associated with KIC 8394721. The position of the vertical dotted lines shows the position of the transits observed for each of the candidates.**

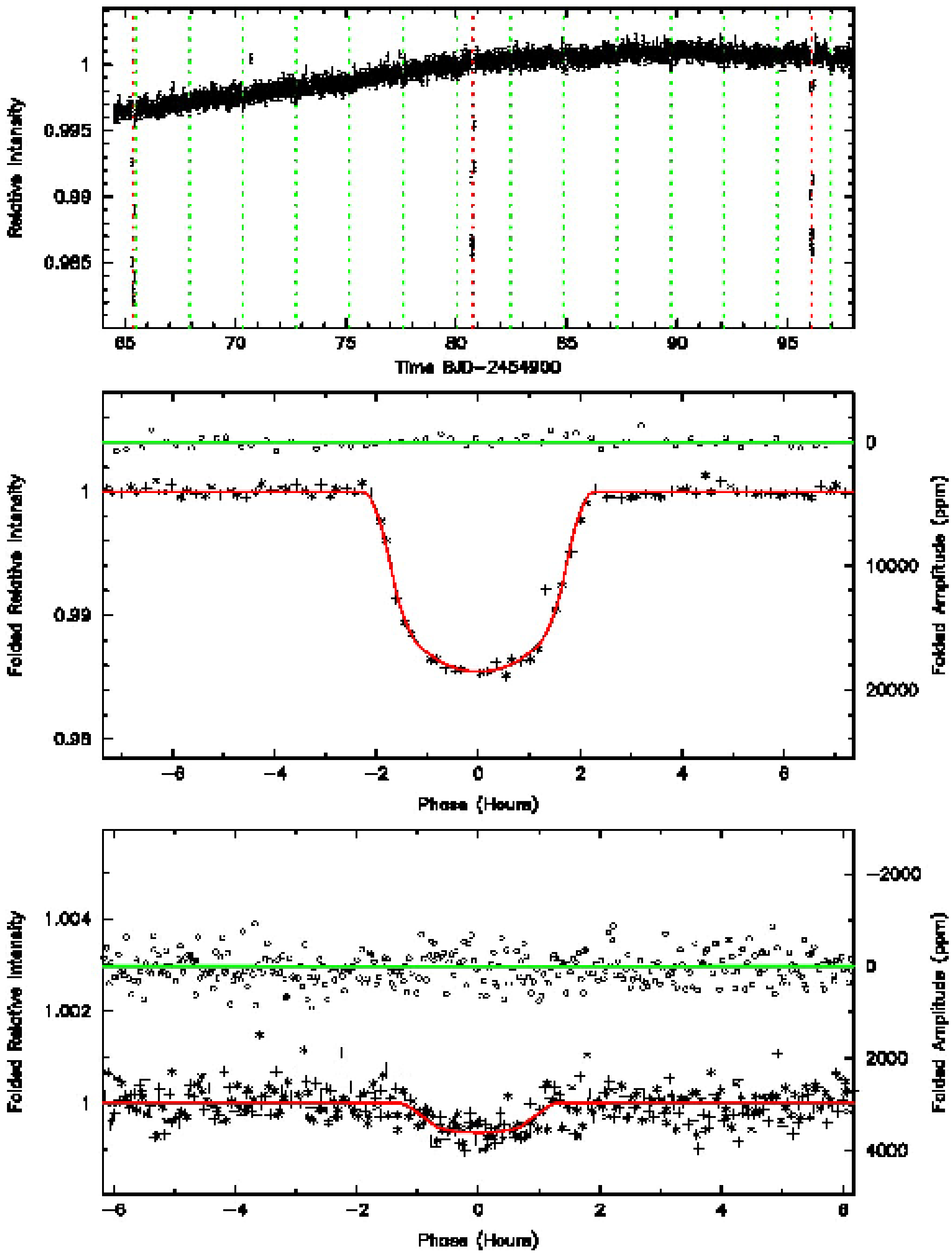

**Figure 10. Two candidate planets associated with KIC 5972334. The clear detection of two candidates (1.06 $R_J$ and 2.0 $R_{\oplus}$) demonstrates that *Kepler* can detect super-Earth-size candidates even for stars as dim as 15$^{th}$ magnitude.**

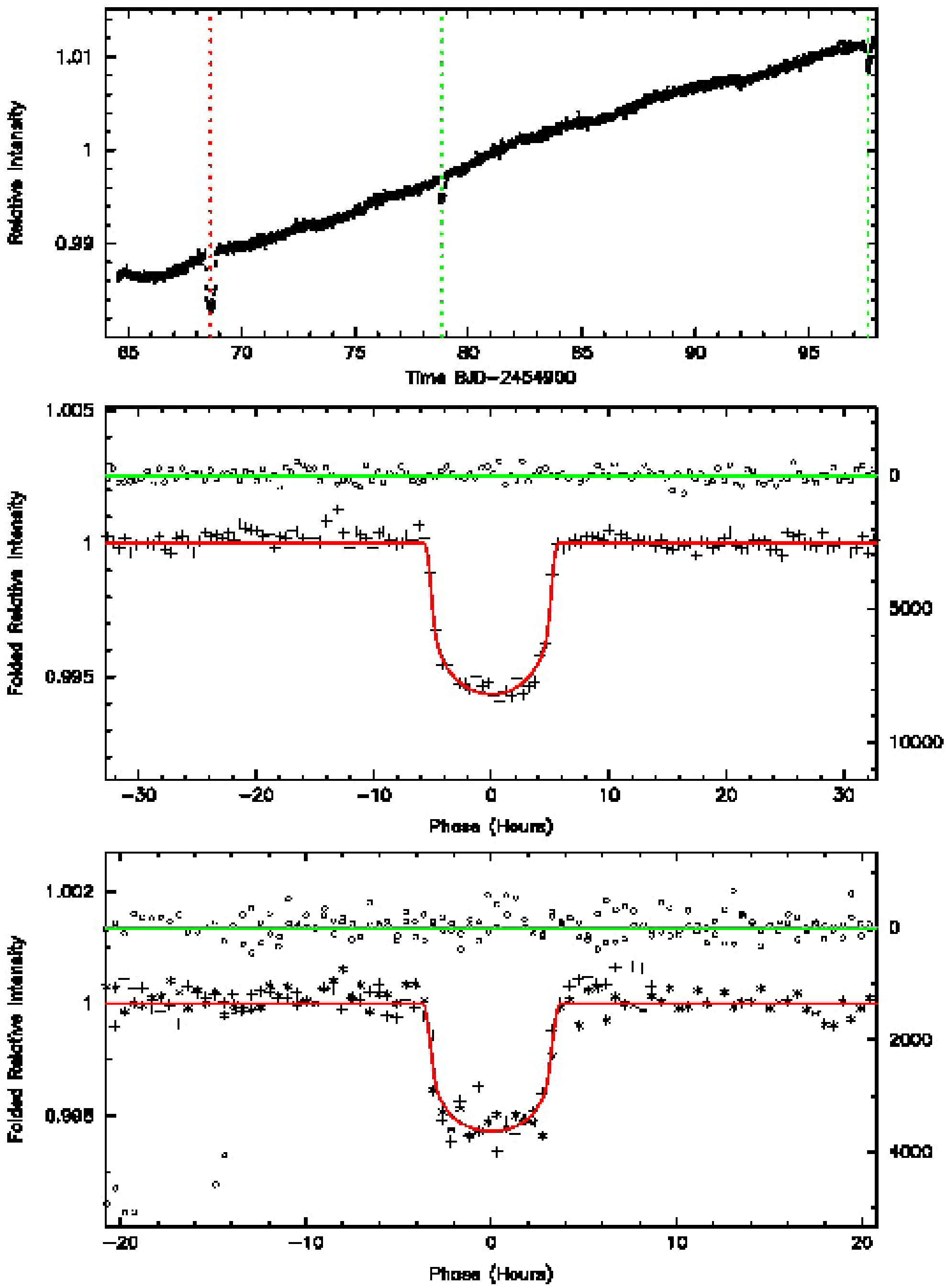

**Figure 11. KIC 10723750. The two sets of transits correspond to two Jupiter-size (1.05 and 0.68 $R_J$) candidates with long periods.**

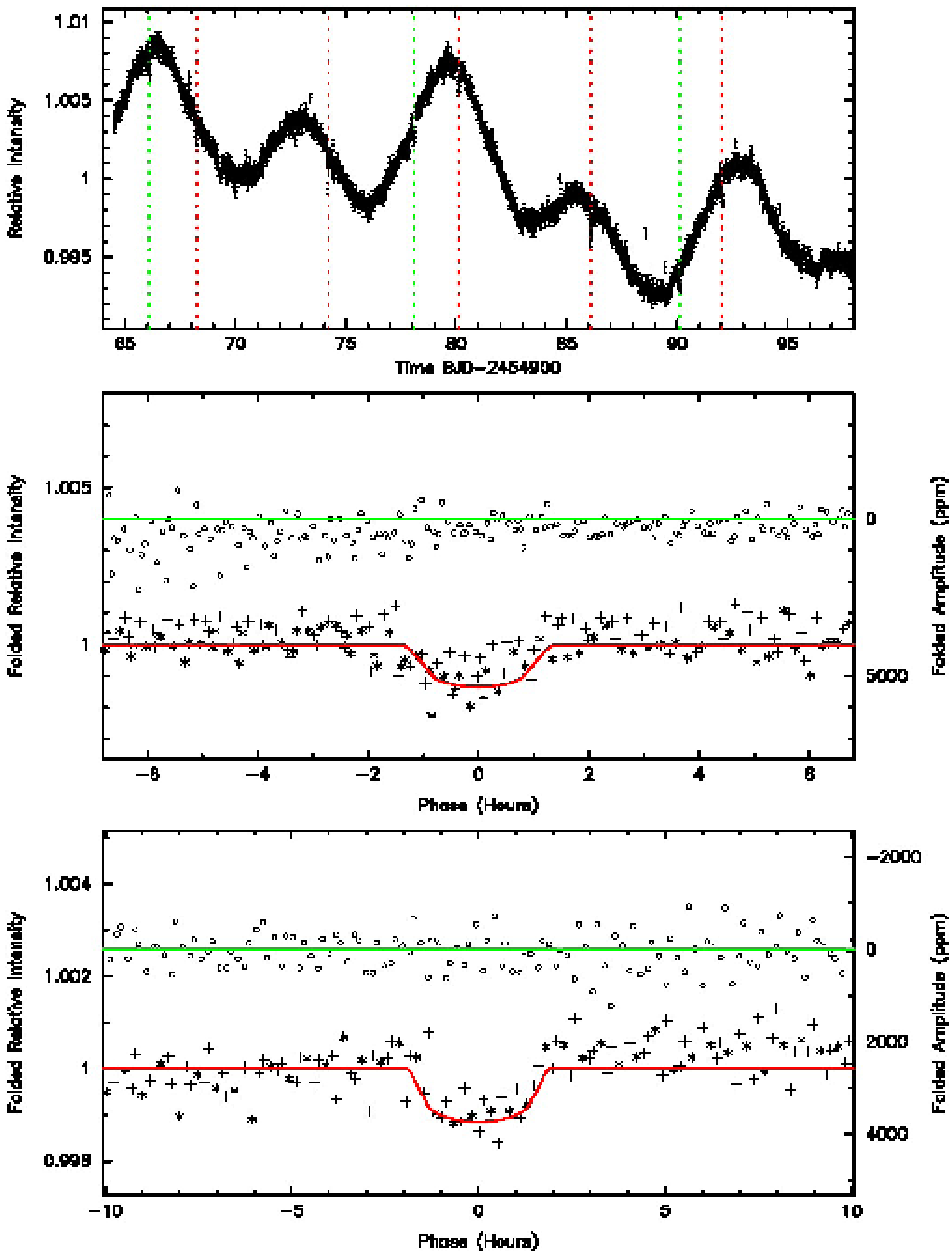

**Figure 12. KIC 7287995. A cool, spotted star with two super-Earth candidates (2.7 and 2.3 R⊕) with near-resonant periods of 5.96 and 12.04 days.**

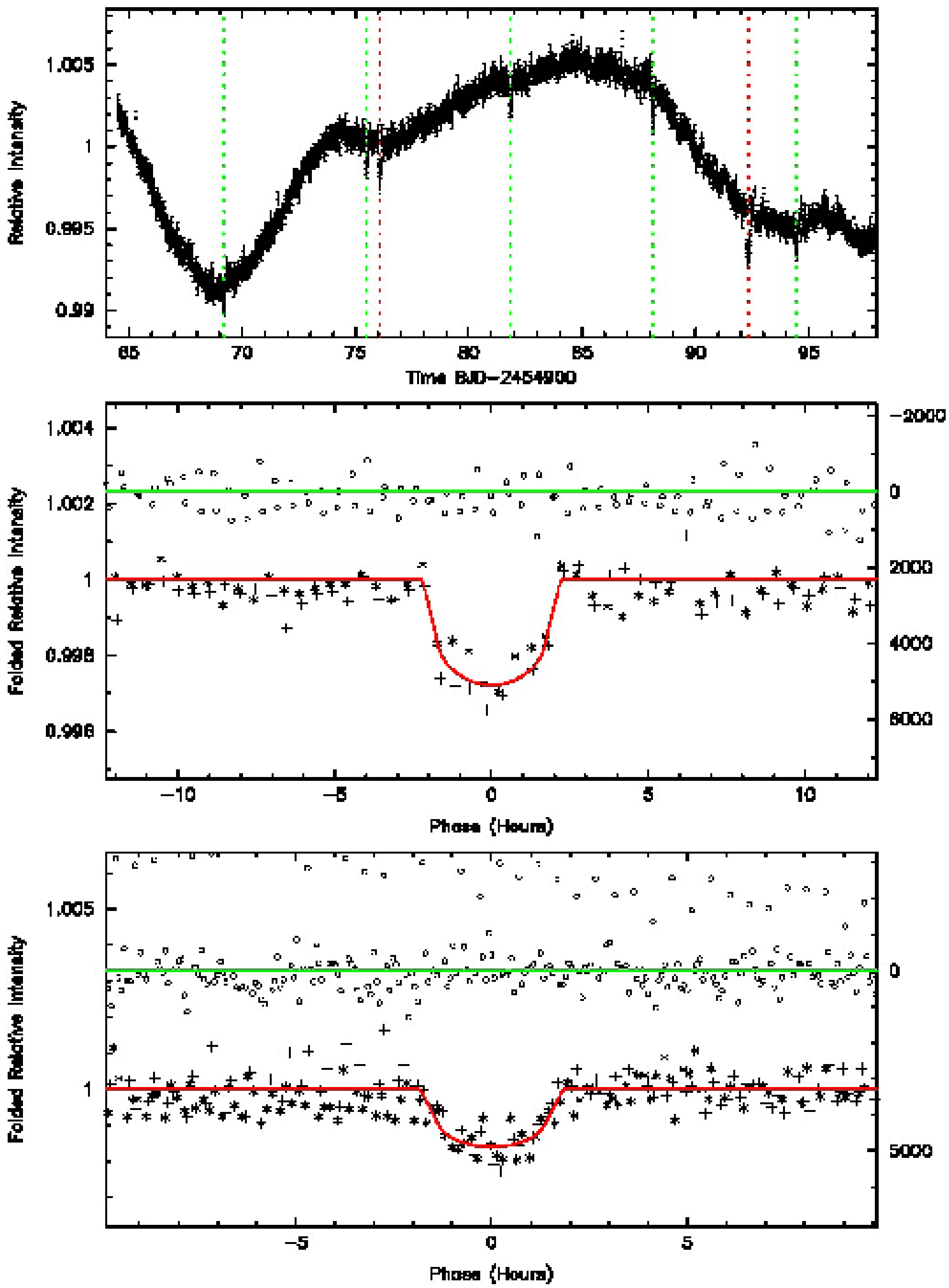

**Figure 13. KIC 7825899. A K-type star with two Neptune-size candidates in 6.3-day and 16.2-day orbits.**

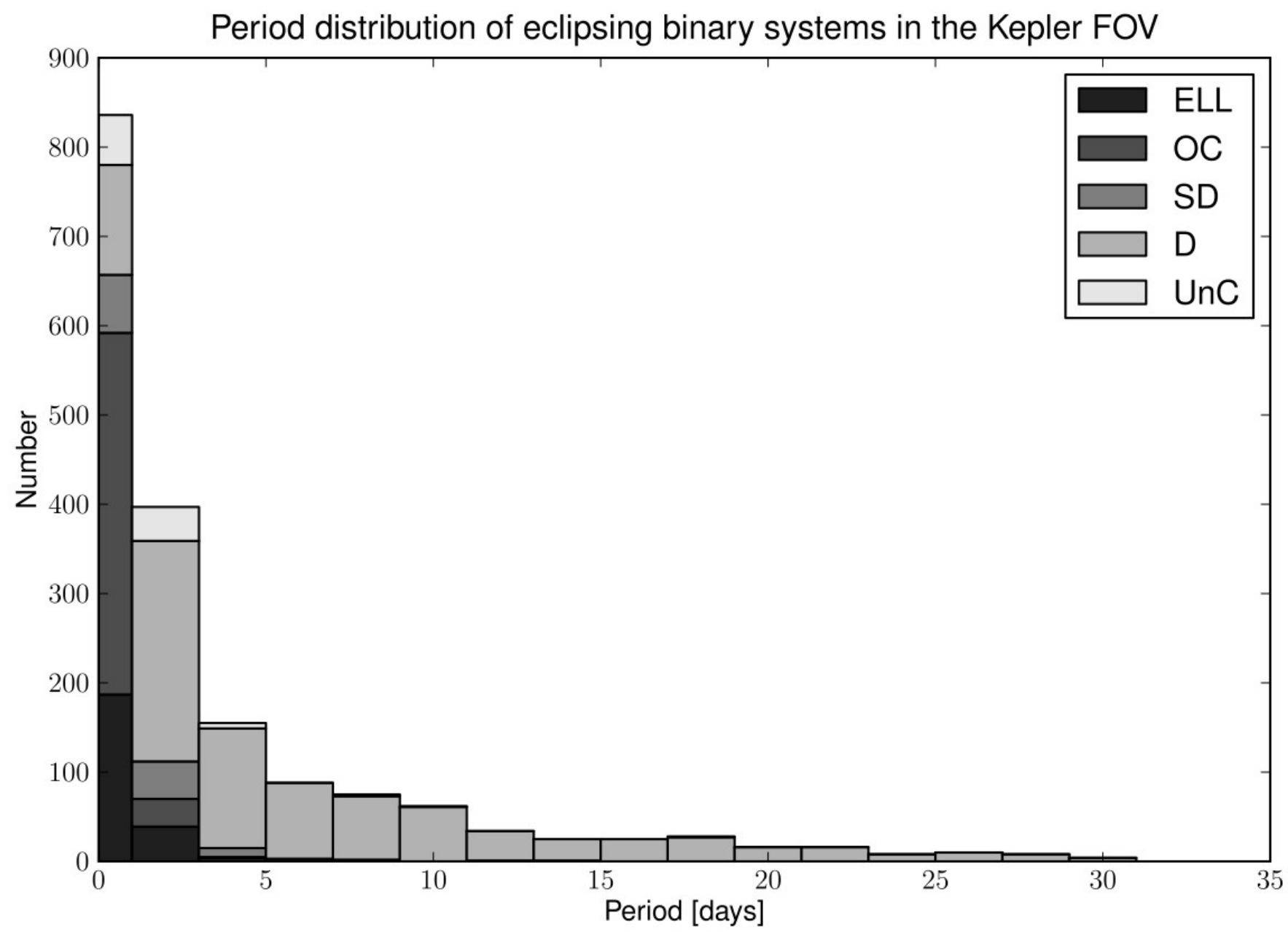

**Figure 14. The histogram of eclipsing binary star periods. Objects are classified into five groups based on their morphologic type: ellipsoidal variables (ELL), overcontact binaries (OC), semi-detached binaries (SD), detached binaries (D) and uncertain (UnC).**

## 5. Eclipsing Binary Data

More than 1.2% of *Kepler* stars are eclipsing binary stars (EBs). Statistical results derived from 1832 EBs are presented by Prsa *et al*. (2010). Figure 14 depicts a distribution of eclipsing binary periods. The stacked gray-scaled bars correspond to different morphologic types. This distribution can be readily compared to that for transiting planets shown in figure 5 for the planetary candidates. The distribution of observed eclipsing binary stars is more heavily weighted towards short periods than is the distribution of planet candidates. This is due to over-contact binaries and ellipsoidal variables, for which there is no counterpart among planets. For a comprehensive discussion of eclipsing binary stars seen in the *Kepler* data, see Prsa *et al*. (2010).

## 6. Summary and Conclusions

The following conclusions must be tempered by recognizing that many sources of bias exist in the results and that the results apply only to the released candidates.

Most candidate planets are less than half the radius of Jupiter.

There is a broad maximum in the frequency of candidates with orbital period in the range from 2 to 5 days. This peak is more prominent for large candidate planets than it is for small candidates.

The observed occurrence frequencies of super-Earth-, Neptune-, Jupiter-, and super-Jupiter-size candidates in short period orbits are approximately $5x10^{-4}$, $3x10^{-3}$, $9x10^{-4}$, and $2x10^{-4}$, respectively. These values are much lower than unbiased values because of no corrections have been made for factors such as stellar size, magnitude, or variability.

The distributions of orbital period and magnitude of the candidates larger than Jupiter appear to be quite different from those of smaller candidates and might represent small stellar companions or errors in the size estimation of the dimmest stars in the *Kepler* planet search program.

One of the five candidate multi-planet systems has two super-Earth-size candidates (2.5 and 2.3 $R_\oplus$) with near-resonant periods of 5.96 and 12.04 days.

**Acknowledgements**

*Kepler* was competitively selected as the tenth Discovery mission. Funding for this mission is provided by NASA's Science Mission Directorate. The authors would like to thank the many people who gave so generously of their time to make this Mission a success.

**Appendix. List of Planetary Candidates**

| KOI | KIC | Kp | Rp (Jupiter) | Epoch BJD-2454900 | Period (days) | Teff (K) | log(g) | R* (Sun) |
|---|---|---|---|---|---|---|---|---|
| 184.01 | 7972785 | 14.9 | 1.59 | 66.566 | 7.301 | 6134 | 4.431 | 1.534 |
| 187.01 | 7023960 | 14.9 | 1.16 | 84.529 | 30.883 | 5768 | 4.703 | 0.829 |
| 188.01 | 5357901 | 14.7 | 0.72 | 66.508 | 3.797 | 5087 | 4.730 | 0.681 |
| 193.01 | 10799735 | 14.9 | 1.46 | 90.349 | 37.590 | 5883 | 4.465 | 1.008 |
| 194.01 | 10904857 | 14.8 | 0.99 | 72.466 | 3.121 | 5883 | 4.633 | 0.820 |
| 195.01 | 11502867 | 14.8 | 1.13 | 66.630 | 3.218 | 5604 | 4.498 | 0.955 |
| 198.01 | 10666242 | 14.3 | 3.43 | 86.369 | 87.233 | 5538 | 4.629 | 0.806 |
| 200.01 | 6046540 | 14.4 | 0.63 | 67.344 | 7.341 | 5774 | 4.690 | 0.759 |
| 204.01 | 9305831 | 14.7 | 0.78 | 66.379 | 3.247 | 5287 | 4.476 | 1.043 |
| 206.01 | 5728139 | 14.5 | 1.20 | 64.982 | 5.334 | 5771 | 4.345 | 1.904 |
| 208.01 | 3762468 | 15.0 | 1.12 | 67.710 | 3.004 | 6094 | 4.585 | 1.176 |
| 210.01 | 10602291 | 14.9 | 1.25 | 72.326 | 20.927 | 5812 | 4.352 | 1.154 |
| 211.01 | 10656508 | 15.0 | 0.94 | 69.014 | 35.875 | 6072 | 4.407 | 1.091 |
| 212.01 | 6300348 | 14.9 | 0.68 | 72.231 | 5.696 | 5843 | 4.538 | 1.056 |
| 214.01 | 11046458 | 14.3 | 1.00 | 64.741 | 3.312 | 5322 | 4.442 | 0.999 |
| 215.01 | 12508335 | 14.7 | 2.70 | 88.206 | 42.944 | 5535 | 4.395 | 1.078 |
| 217.01 | 9595827 | 15.1 | 1.18 | 66.414 | 3.905 | 5504 | 4.724 | 0.896 |
| 219.01 | 6305192 | 14.2 | 0.68 | 65.470 | 8.025 | 5347 | 4.727 | 1.372 |
| 220.01 | 7132798 | 14.2 | 0.38 | 65.939 | 2.422 | 5388 | 4.867 | 0.989 |
| 221.01 | 3937519 | 14.6 | 0.49 | 65.441 | 3.413 | 5176 | 4.686 | 0.898 |
| 223.01 | 4545187 | 14.7 | 0.24 | 67.478 | 3.177 | 5128 | 4.657 | 0.744 |
| 224.01 | 5547480 | 14.8 | 0.32 | 65.073 | 3.980 | 5740 | 4.507 | 0.951 |
| 225.01 | 5801571 | 14.8 | 0.45 | 74.537 | 0.839 | 6037 | 4.546 | 0.919 |
| 226.01 | 5959753 | 14.8 | 0.22 | 71.116 | 8.309 | 5043 | 4.892 | 0.869 |
| 229.01 | 3847907 | 14.7 | 0.56 | 67.934 | 3.573 | 5608 | 4.370 | 1.119 |
| 234.01 | 8491277 | 14.3 | 0.29 | 65.187 | 9.614 | 5735 | 4.356 | 1.205 |
| 235.01 | 8107225 | 14.4 | 0.18 | 66.818 | 5.632 | 5041 | 4.654 | 0.740 |
| 237.01 | 8041216 | 14.2 | 0.20 | 67.788 | 8.508 | 5679 | 4.533 | 0.919 |
| 239.01 | 6383785 | 14.8 | 0.31 | 71.556 | 5.641 | 5983 | 4.539 | 0.924 |
| 240.01 | 8026752 | 15.0 | 0.45 | 71.615 | 4.287 | 5996 | 4.602 | 1.446 |
| 241.01 | 11288051 | 14.1 | 0.19 | 64.796 | 13.821 | 5055 | 4.854 | 0.689 |
| 242.01 | 3642741 | 14.7 | 0.86 | 71.343 | 7.259 | 5437 | 4.507 | 1.556 |
| 403.01 | 4247092 | 14.2 | 1.58 | 104.132 | 21.057 | 5565 | 4.440 | 1.022 |
| 409.01 | 5444548 | 14.2 | 0.21 | 112.522 | 13.249 | 5709 | 5.008 | 0.993 |
| 410.01 | 5449777 | 14.5 | 1.07 | 109.286 | 7.217 | 5968 | 4.384 | 1.117 |
| 412.01 | 5683743 | 14.3 | 0.72 | 103.325 | 4.147 | 5584 | 4.275 | 1.256 |
| 413.01 | 5791986 | 14.8 | 0.28 | 109.558 | 15.229 | 5236 | 4.557 | 8.560 |
| 416.01 | 6508221 | 14.3 | 0.27 | 118.841 | 18.208 | 5083 | 4.647 | 0.750 |
| 417.01 | 6879865 | 14.8 | 0.81 | 109.965 | 19.193 | 5635 | 4.594 | 0.851 |
| 418.01 | 7975727 | 14.5 | 1.03 | 105.796 | 22.418 | 5153 | 4.422 | 1.010 |
| 419.01 | 8219673 | 14.5 | 0.67 | 122.391 | 20.131 | 5723 | 4.695 | 0.752 |
| 420.01 | 8352537 | 14.2 | 0.42 | 107.084 | 6.010 | 4687 | 4.513 | 0.831 |
| 421.01 | 9115800 | 15.0 | 1.60 | 105.819 | 4.454 | 5181 | 4.317 | 1.158 |
| 423.01 | 9478990 | 14.3 | 0.94 | 135.857 | 21.087 | 5992 | 4.448 | 1.138 |
| 425.01 | 9967884 | 14.7 | 0.43 | 102.753 | 5.428 | 5689 | 4.544 | 0.438 |
| 426.01 | 10016874 | 14.7 | 0.36 | 105.152 | 16.301 | 5796 | 4.328 | 1.188 |
| 427.01 | 10189546 | 14.6 | 0.43 | 124.737 | 24.615 | 5293 | 4.496 | 0.930 |
| 428.01 | 10418224 | 14.6 | 1.04 | 105.518 | 6.873 | 6127 | 4.549 | 1.927 |
| 429.01 | 10616679 | 14.5 | 0.48 | 105.527 | 8.600 | 5093 | 4.485 | 1.024 |
| 430.01 | 10717241 | 14.9 | 0.25 | 112.402 | 12.377 | 4124 | 4.584 | 0.640 |
| 431.01 | 10843590 | 14.3 | 0.34 | 111.712 | 18.870 | 5249 | 4.433 | 1.004 |
| 432.01 | 10858832 | 14.3 | 0.32 | 107.350 | 5.263 | 5830 | 4.457 | 1.015 |
| 433.01 | 10937029 | 14.9 | 0.52 | 104.095 | 4.030 | 5237 | 4.372 | 1.084 |
| 434.01 | 11656302 | 14.6 | 1.69 | 106.103 | 22.265 | 5172 | 4.564 | 1.350 |
| 435.01 | 11709124 | 14.5 | 0.36 | 111.951 | 20.548 | 5709 | 4.663 | 1.039 |
| 438.01 | 12302530 | 14.3 | 0.19 | 107.796 | 5.931 | 4351 | 4.595 | 0.679 |
| 441.01 | 3340312 | 14.5 | 0.24 | 106.917 | 30.544 | 6231 | 4.628 | 0.838 |
| 443.01 | 3833007 | 14.2 | 0.24 | 113.046 | 16.217 | 5614 | 4.617 | 1.020 |
| 450.01 | 6042214 | 14.2 | 0.30 | 104.953 | 27.047 | 6089 | 4.561 | 0.904 |
| 451.01 | 6200715 | 14.9 | 0.23 | 105.178 | 3.724 | 6333 | 4.648 | 1.012 |
| 452.01 | 6291033 | 14.6 | 0.34 | 102.939 | 3.706 | 5935 | 4.409 | 1.771 |
| 454.01 | 7098355 | 14.8 | 0.21 | 103.557 | 29.007 | 5138 | 4.569 | 0.835 |
| 456.01 | 7269974 | 14.6 | 0.27 | 104.476 | 13.700 | 5644 | 4.515 | 0.950 |
| 457.01 | 7440748 | 14.2 | 0.19 | 107.295 | 4.921 | 4931 | 4.650 | 0.729 |
| 458.01 | 7504328 | 14.7 | 0.94 | 141.081 | 53.717 | 5593 | 4.280 | 1.248 |
| 459.01 | 7977197 | 14.2 | 0.32 | 103.102 | 19.447 | 5601 | 4.428 | 1.040 |
| 460.01 | 8043638 | 14.7 | 0.41 | 109.077 | 17.587 | 5387 | 4.334 | 1.150 |
| 466.01 | 9008220 | 14.7 | 0.28 | 103.538 | 9.391 | 5907 | 4.896 | 0.590 |
| 467.01 | 9583881 | 14.8 | 0.48 | 115.442 | 18.009 | 5583 | 4.539 | 0.979 |
| 468.01 | 9589524 | 14.8 | 0.36 | 107.596 | 22.184 | 4999 | 4.499 | 0.900 |
| 469.01 | 9703198 | 14.7 | 0.49 | 107.607 | 10.329 | 6005 | 4.631 | 0.827 |
| 470.01 | 9844088 | 14.7 | 0.35 | 104.150 | 3.751 | 5542 | 4.653 | 0.782 |
| 471.01 | 10019643 | 14.4 | 0.17 | 104.730 | 21.348 | 5548 | 4.670 | 0.766 |
| 472.01 | 10123064 | 15.0 | 0.38 | 106.565 | 4.244 | 5682 | 4.580 | 1.149 |
| 473.01 | 10155434 | 14.7 | 0.22 | 113.637 | 12.705 | 5379 | 4.686 | 0.737 |
| 474.01 | 10460984 | 14.3 | 0.22 | 109.721 | 10.946 | 6143 | 4.468 | 1.015 |
| 476.01 | 10599206 | 15.0 | 0.23 | 111.437 | 18.428 | 4993 | 4.514 | 0.881 |
| 477.01 | 10934674 | 14.7 | 0.23 | 102.646 | 16.542 | 5039 | 4.513 | 0.889 |
| 480.01 | 11134879 | 14.3 | 0.24 | 105.308 | 4.302 | 5324 | 4.511 | 0.915 |
| 482.01 | 11255761 | 14.9 | 0.30 | 102.552 | 4.993 | 5526 | 4.426 | 1.036 |
| 483.01 | 11497977 | 14.7 | 0.23 | 106.257 | 4.799 | 5410 | 4.703 | 0.938 |
| 484.01 | 12061222 | 14.5 | 0.20 | 108.064 | 17.204 | 5065 | 4.759 | 0.745 |
| 486.01 | 12404305 | 14.1 | 0.22 | 102.492 | 22.184 | 5625 | 5.000 | 0.969 |
| 487.01 | 12834874 | 14.5 | 0.17 | 106.036 | 7.659 | 5463 | 4.510 | 0.977 |
| 488.01 | 2557816 | 14.7 | 0.18 | 109.444 | 9.380 | 5488 | 4.490 | 0.955 |
| 491.01 | 3541800 | 14.4 | 0.15 | 102.670 | 4.662 | 5965 | 4.684 | 0.798 |
| 492.01 | 3559935 | 14.4 | 0.32 | 127.712 | 29.910 | 5373 | 4.263 | 1.258 |
| 493.01 | 3834360 | 14.7 | 0.21 | 103.125 | 2.908 | 5583 | 4.571 | 0.871 |
| 494.01 | 3966801 | 14.9 | 0.17 | 121.780 | 25.698 | 4854 | 4.904 | 0.620 |
| 497.01 | 4757437 | 14.6 | 0.24 | 108.609 | 13.193 | 6045 | 4.495 | 1.163 |
| 499.01 | 4847534 | 14.3 | 0.17 | 107.535 | 9.669 | 5362 | 4.531 | 0.896 |
| 501.01 | 4951877 | 14.6 | 0.29 | 103.340 | 24.793 | 5556 | 4.501 | 1.502 |
| 502.01 | 5282051 | 14.3 | 0.18 | 104.159 | 5.910 | 5288 | 4.339 | 1.134 |
| 503.01 | 5340644 | 15.0 | 0.23 | 105.958 | 8.222 | 4110 | 4.550 | 0.673 |
| 504.01 | 5461440 | 14.6 | 0.17 | 132.291 | 40.588 | 5403 | 4.754 | 0.678 |
| 505.01 | 5689351 | 14.2 | 0.33 | 107.812 | 13.767 | 4985 | 4.242 | 1.259 |
| 506.01 | 5780715 | 14.7 | 0.25 | 102.966 | 1.583 | 5777 | 4.557 | 0.896 |
| 507.01 | 5812960 | 14.9 | 0.41 | 106.494 | 18.495 | 5117 | 4.408 | 1.024 |
| 509.01 | 6381846 | 14.9 | 0.23 | 102.712 | 4.167 | 5437 | 4.565 | 0.900 |
| 511.01 | 6451936 | 14.2 | 0.25 | 103.504 | 8.006 | 5802 | 4.404 | 1.083 |
| 512.01 | 6838050 | 14.8 | 0.25 | 105.919 | 6.510 | 5406 | 4.316 | 1.178 |
| 513.01 | 6937692 | 14.9 | 0.30 | 103.098 | 35.181 | 6288 | 4.577 | 1.204 |
| 514.01 | 7602070 | 14.4 | 0.17 | 109.061 | 11.757 | 5446 | 4.916 | 0.841 |
| 519.01 | 8022244 | 14.9 | 0.21 | 111.337 | 11.904 | 5807 | 4.523 | 0.991 |
| 520.01 | 8037145 | 14.6 | 0.26 | 103.304 | 12.760 | 5048 | 4.465 | 0.946 |
| 521.01 | 8162789 | 14.6 | 0.41 | 105.003 | 10.161 | 5767 | 4.394 | 1.094 |
| 522.01 | 8265218 | 14.4 | 0.19 | 102.940 | 12.831 | 5663 | 4.910 | 0.631 |
| 523.01 | 8806123 | 15.0 | 0.63 | 131.230 | 49.413 | 5942 | 4.421 | 1.066 |
| 524.01 | 8934495 | 14.9 | 0.20 | 104.997 | 4.593 | 5187 | 4.698 | 0.720 |
| 525.01 | 9119458 | 14.5 | 0.49 | 106.678 | 11.532 | 5524 | 4.281 | 1.241 |
| 526.01 | 9157634 | 14.4 | 0.22 | 104.044 | 2.105 | 5467 | 4.633 | 0.796 |
| 528.01 | 9941859 | 14.6 | 0.29 | 109.674 | 9.577 | 5448 | 4.346 | 1.138 |
| 532.01 | 10454313 | 14.7 | 0.23 | 106.689 | 4.222 | 5874 | 4.540 | 1.033 |
| 533.01 | 10513530 | 14.7 | 0.22 | 104.698 | 16.550 | 5198 | 4.444 | 0.985 |
| 535.01 | 10873260 | 14.4 | 0.40 | 104.182 | 5.853 | 5782 | 4.450 | 1.358 |
| 537.01 | 11073351 | 14.7 | 0.18 | 103.785 | 2.820 | 5889 | 4.906 | 0.949 |
| 538.01 | 11090765 | 14.6 | 0.24 | 104.657 | 21.214 | 5923 | 4.427 | 1.061 |
| 539.01 | 11246364 | 14.1 | 0.15 | 104.196 | 29.122 | 5722 | 4.361 | 1.137 |
| 540.01 | 11521048 | 14.9 | 0.74 | 127.824 | 25.703 | 5361 | 4.498 | 0.934 |
| 541.01 | 11656721 | 14.7 | 0.15 | 113.347 | 13.647 | 5369 | 4.712 | 0.741 |
| 542.01 | 11669239 | 14.3 | 0.24 | 111.682 | 41.889 | 5509 | 4.357 | 1.128 |
| 543.01 | 11823054 | 14.7 | 0.17 | 106.438 | 4.302 | 5166 | 4.724 | 0.686 |
| 544.01 | 11913012 | 14.8 | 0.15 | 104.669 | 3.748 | 5883 | 4.585 | 1.012 |
| 546.01 | 12058931 | 14.9 | 0.31 | 103.186 | 20.686 | 5989 | 4.487 | 1.244 |
| 547.01 | 12116489 | 14.8 | 0.32 | 121.061 | 25.302 | 5086 | 4.619 | 0.788 |
| 549.01 | 3437776 | 14.6 | 0.44 | 126.512 | 42.895 | 5609 | 4.414 | 1.059 |
| 551.01 | 4270253 | 14.9 | 0.17 | 111.850 | 11.636 | 5627 | 4.667 | 0.775 |
| 552.01 | 5122112 | 14.7 | 1.00 | 104.099 | 3.055 | 6018 | 4.431 | 1.057 |
| 553.01 | 5303551 | 14.9 | 0.20 | 104.453 | 2.399 | 5404 | 4.394 | 1.140 |
| 554.01 | 5443837 | 14.5 | 0.39 | 103.544 | 3.658 | 5835 | 4.641 | 0.809 |
| 557.01 | 5774349 | 15.0 | 0.28 | 103.785 | 15.656 | 5002 | 4.415 | 1.005 |
| 558.01 | 5978361 | 14.9 | 0.22 | 106.084 | 9.179 | 5281 | 4.580 | 0.835 |
| 559.01 | 6422367 | 14.8 | 0.17 | 106.712 | 4.330 | 5187 | 4.467 | 0.955 |
| 560.01 | 6501635 | 14.7 | 0.16 | 112.266 | 23.678 | 5142 | 4.834 | 0.750 |
| 563.01 | 6707833 | 14.5 | 0.18 | 108.632 | 15.284 | 5879 | 4.477 | 1.173 |
| 564.01 | 6786037 | 14.9 | 0.31 | 104.887 | 21.060 | 5686 | 4.525 | 1.453 |
| 565.01 | 7025846 | 14.3 | 0.14 | 103.202 | 2.340 | 5829 | 4.409 | 1.068 |
| 569.01 | 8008206 | 14.5 | 0.21 | 118.442 | 20.725 | 5039 | 4.546 | 0.851 |
| 570.01 | 8106610 | 14.8 | 0.27 | 105.782 | 12.399 | 6079 | 4.452 | 1.033 |
| 572.01 | 8193178 | 14.2 | 0.23 | 112.777 | 10.640 | 5666 | 4.310 | 1.325 |
| 573.01 | 8344004 | 14.7 | 0.28 | 105.505 | 5.997 | 5729 | 4.352 | 1.149 |
| 574.01 | 8355239 | 14.9 | 0.21 | 104.362 | 20.136 | 5047 | 4.669 | 0.727 |
| 575.01 | 8367113 | 14.7 | 0.23 | 116.405 | 24.321 | 5979 | 4.480 | 0.994 |
| 578.01 | 8565266 | 14.7 | 0.46 | 102.879 | 6.413 | 5777 | 4.362 | 1.528 |
| 580.01 | 8625925 | 14.9 | 0.19 | 108.711 | 6.521 | 5603 | 4.920 | 0.806 |
| 581.01 | 8822216 | 14.8 | 0.23 | 108.914 | 6.997 | 5514 | 4.856 | 0.761 |
| 582.01 | 9020160 | 14.8 | 0.20 | 103.467 | 5.945 | 5103 | 4.650 | 0.750 |
| 583.01 | 9076513 | 14.6 | 0.17 | 103.740 | 2.437 | 5735 | 4.550 | 1.197 |
| 585.01 | 9279669 | 14.9 | 0.18 | 104.558 | 3.722 | 5437 | 4.737 | 0.695 |
| 586.01 | 9570741 | 14.6 | 0.16 | 108.979 | 15.779 | 5707 | 4.669 | 0.802 |
| 587.01 | 9607164 | 14.6 | 0.28 | 104.606 | 14.034 | 5112 | 4.423 | 1.005 |
| 588.01 | 9631762 | 14.3 | 0.21 | 108.672 | 10.356 | 4431 | 4.459 | 0.852 |
| 590.01 | 9782691 | 14.6 | 0.16 | 107.545 | 11.389 | 6106 | 4.546 | 0.922 |
| 592.01 | 9957627 | 14.3 | 0.24 | 108.475 | 39.759 | 5810 | 4.408 | 1.077 |
| 593.01 | 9958962 | 15.0 | 0.19 | 104.792 | 9.997 | 5737 | 4.617 | 0.889 |
| 597.01 | 10600261 | 14.9 | 0.20 | 109.942 | 17.308 | 5833 | 4.416 | 1.046 |
| 598.01 | 10656823 | 14.8 | 0.18 | 104.152 | 8.309 | 5171 | 4.811 | 0.749 |
| 599.01 | 10676824 | 14.9 | 0.20 | 106.212 | 6.455 | 5820 | 4.540 | 0.916 |
| 600.01 | 10718726 | 14.8 | 0.18 | 103.367 | 3.596 | 5869 | 4.445 | 1.032 |
| 602.01 | 12459913 | 14.6 | 0.23 | 110.276 | 12.914 | 6007 | 4.405 | 1.282 |
| 605.01 | 4832837 | 14.9 | 0.16 | 102.718 | 2.628 | 4270 | 4.757 | 0.581 |
| 607.01 | 5441980 | 14.4 | 0.57 | 106.492 | 5.894 | 5497 | 4.608 | 0.825 |
| 608.01 | 5562784 | 14.7 | 0.47 | 125.921 | 25.333 | 4324 | 4.551 | 1.326 |
| 609.01 | 5608566 | 14.5 | 1.20 | 105.027 | 4.397 | 5696 | 4.295 | 1.231 |
| 610.01 | 5686174 | 14.7 | 0.19 | 113.850 | 14.281 | 4072 | 4.529 | 0.687 |
| 614.01 | 7368664 | 14.5 | 0.36 | 103.023 | 12.875 | 5675 | 4.887 | 0.589 |
| 617.01 | 9846086 | 14.6 | 2.06 | 131.599 | 37.865 | 5594 | 4.530 | 0.917 |
| 618.01 | 10353968 | 15.0 | 0.28 | 111.347 | 9.071 | 5471 | 4.516 | 0.922 |
| 620.01 | 11773022 | 14.7 | 0.65 | 92.107 | 45.154 | 5803 | 4.544 | 1.384 |
| 725.01 | 10068383 | 15.8 | 0.75 | 102.644 | 7.305 | 5046 | 4.652 | 0.882 |
| 726.01 | 10157573 | 15.1 | 0.30 | 106.266 | 5.116 | 6164 | 4.508 | 0.969 |
| 728.01 | 10221013 | 15.4 | 0.89 | 103.121 | 7.189 | 5976 | 4.544 | 0.918 |
| 729.01 | 10225800 | 15.6 | 0.36 | 102.674 | 1.424 | 5707 | 4.608 | 0.838 |
| 730.01 | 10227020 | 15.3 | 0.31 | 109.793 | 14.785 | 5599 | 4.386 | 1.287 |
| 732.01 | 10265898 | 15.3 | 0.25 | 103.407 | 1.260 | 5360 | 4.588 | 0.860 |
| 733.01 | 10271806 | 15.6 | 0.24 | 102.725 | 5.925 | 5038 | 4.846 | 0.730 |
| 734.01 | 10272442 | 15.3 | 0.39 | 120.924 | 24.542 | 5719 | 4.700 | 1.329 |
| 736.01 | 10340423 | 16.0 | 0.25 | 110.789 | 18.796 | 4157 | 4.552 | 0.681 |
| 737.01 | 10345478 | 15.7 | 0.43 | 115.678 | 14.499 | 5117 | 4.602 | 0.798 |
| 740.01 | 10395381 | 15.6 | 0.17 | 119.368 | 17.672 | 4711 | 4.640 | 0.703 |
| 743.01 | 10464078 | 15.5 | 1.65 | 105.491 | 19.402 | 4877 | 4.304 | 1.904 |
| 746.01 | 10526549 | 15.3 | 0.24 | 106.246 | 9.274 | 4681 | 4.551 | 0.788 |
| 747.01 | 10583066 | 15.8 | 0.28 | 104.602 | 6.029 | 4357 | 4.680 | 0.608 |
| 749.01 | 10601284 | 15.4 | 0.23 | 104.806 | 5.350 | 5374 | 4.780 | 0.915 |
| 750.01 | 10662202 | 15.4 | 0.19 | 104.535 | 21.679 | 4619 | 4.624 | 0.703 |
| 752.01 | 10797460 | 15.3 | 0.26 | 103.533 | 9.489 | 5584 | 4.406 | 1.067 |
| 753.01 | 10811496 | 15.4 | 2.11 | 108.840 | 19.904 | 5648 | 4.843 | 0.621 |
| 758.01 | 10987985 | 15.4 | 0.43 | 109.353 | 16.016 | 4869 | 4.284 | 1.172 |
| 759.01 | 11018648 | 15.1 | 0.32 | 127.134 | 32.629 | 5401 | 4.563 | 0.864 |
| 760.01 | 11138155 | 15.3 | 0.81 | 105.257 | 4.959 | 5887 | 4.622 | 0.830 |
| 762.01 | 11153539 | 15.4 | 0.24 | 104.356 | 4.498 | 5779 | 4.596 | 1.172 |
| 764.01 | 11304958 | 15.4 | 0.71 | 141.932 | 41.441 | 5263 | 4.367 | 1.582 |
| 765.01 | 11391957 | 15.3 | 0.20 | 104.629 | 8.354 | 5345 | 4.700 | 0.722 |
| 769.01 | 11460018 | 15.4 | 0.21 | 104.903 | 4.281 | 5461 | 4.643 | 0.942 |
| 770.01 | 11463211 | 15.5 | 0.26 | 103.998 | 1.506 | 5502 | 4.927 | 0.590 |
| 771.01 | 11465813 | 15.2 | 1.34 | 142.039 | 10389.109 | 5574 | 4.380 | 9.058 |
| 772.01 | 11493732 | 15.2 | 0.68 | 106.831 | 61.263 | 5885 | 4.409 | 1.079 |
| 773.01 | 11507101 | 15.2 | 0.21 | 105.837 | 38.374 | 5667 | 4.624 | 0.820 |
| 776.01 | 11812062 | 15.5 | 0.55 | 104.792 | 3.729 | 5309 | 4.829 | 0.843 |
| 777.01 | 11818800 | 15.5 | 2.00 | 106.564 | 40.420 | 5256 | 4.479 | 0.948 |
| 778.01 | 11853255 | 15.1 | 0.18 | 103.681 | 2.243 | 4082 | 4.605 | 0.611 |
| 782.01 | 11960862 | 15.3 | 0.59 | 106.634 | 6.575 | 5733 | 4.411 | 1.248 |
| 783.01 | 12020329 | 15.1 | 0.96 | 102.991 | 7.275 | 5284 | 4.762 | 1.953 |
| 784.01 | 12066335 | 15.4 | 0.25 | 119.798 | 19.266 | 4112 | 4.569 | 0.653 |
| 785.01 | 12070811 | 15.5 | 0.21 | 111.749 | 12.393 | 5380 | 4.725 | 0.741 |
| 786.01 | 12110942 | 15.2 | 0.17 | 103.366 | 3.690 | 5638 | 4.715 | 0.876 |
| 787.01 | 12366084 | 15.4 | 0.30 | 104.017 | 4.431 | 5615 | 4.534 | 1.037 |
| 788.01 | 12404086 | 15.2 | 0.31 | 109.049 | 26.396 | 4950 | 4.634 | 0.747 |
| 789.01 | 12459725 | 15.7 | 0.15 | 104.505 | 14.180 | 5563 | 4.765 | 0.683 |
| 790.01 | 12470844 | 15.3 | 0.18 | 107.168 | 8.472 | 5176 | 5.058 | 0.612 |
| 791.01 | 12644822 | 15.1 | 0.79 | 113.890 | 12.612 | 5564 | 4.528 | 1.117 |
| 793.01 | 2445129 | 15.1 | 0.34 | 106.313 | 10.319 | 5655 | 4.409 | 1.069 |
| 795.01 | 3114167 | 15.6 | 0.22 | 103.575 | 6.770 | 5455 | 4.804 | 0.640 |
| 799.01 | 3246984 | 15.3 | 0.41 | 102.817 | 1.627 | 5491 | 4.412 | 1.051 |
| 802.01 | 3453214 | 15.6 | 0.75 | 114.881 | 19.620 | 5556 | 5.009 | 0.498 |
| 804.01 | 3641726 | 15.4 | 0.23 | 110.194 | 9.030 | 5136 | 4.533 | 0.874 |
| 808.01 | 3838486 | 15.8 | 0.37 | 104.985 | 2.990 | 4389 | 4.582 | 0.701 |
| 810.01 | 3940418 | 15.1 | 0.23 | 103.507 | 4.783 | 4997 | 4.571 | 0.820 |
| 811.01 | 4049131 | 15.4 | 0.41 | 114.427 | 20.507 | 4764 | 4.432 | 0.944 |
| 812.01 | 4139816 | 16.0 | 0.22 | 104.978 | 3.340 | 4097 | 4.661 | 0.571 |
| 813.01 | 4275191 | 15.7 | 0.60 | 103.528 | 3.896 | 5357 | 4.726 | 0.725 |
| 814.01 | 4476123 | 15.6 | 0.27 | 108.450 | 22.368 | 5236 | 4.855 | 0.984 |
| 815.01 | 4544670 | 15.7 | 0.90 | 105.628 | 34.845 | 5344 | 4.485 | 0.948 |
| 819.01 | 4932348 | 15.5 | 1.39 | 129.933 | 38.037 | 5386 | 4.963 | 0.518 |
| 820.01 | 4936180 | 15.3 | 0.67 | 106.720 | 4.641 | 6287 | 4.511 | 0.970 |
| 822.01 | 5077629 | 15.8 | 0.95 | 105.179 | 7.919 | 5458 | 4.605 | 0.824 |
| 823.01 | 5115978 | 15.2 | 0.82 | 103.228 | 1.028 | 5976 | 4.427 | 4.223 |
| 825.01 | 5252423 | 15.3 | 0.19 | 109.957 | 8.103 | 4735 | 4.581 | 0.764 |
| 826.01 | 5272878 | 15.1 | 0.21 | 104.135 | 6.366 | 5557 | 4.843 | 0.854 |
| 827.01 | 5283542 | 15.5 | 0.25 | 107.779 | 5.975 | 5837 | 4.539 | 0.918 |
| 829.01 | 5358241 | 15.4 | 0.24 | 107.778 | 18.649 | 5858 | 4.567 | 0.888 |
| 833.01 | 5376067 | 15.4 | 0.39 | 106.275 | 3.951 | 5781 | 4.660 | 0.788 |
| 834.01 | 5436502 | 15.1 | 0.78 | 104.372 | 23.655 | 5614 | 4.598 | 1.496 |
| 835.01 | 5456651 | 15.2 | 0.17 | 113.936 | 11.763 | 4817 | 4.952 | 0.635 |
| 837.01 | 5531576 | 15.7 | 0.16 | 107.659 | 7.954 | 4817 | 4.751 | 0.623 |
| 838.01 | 5534814 | 15.3 | 0.69 | 106.011 | 4.859 | 5794 | 4.475 | 0.991 |
| 842.01 | 5794379 | 15.4 | 0.25 | 108.349 | 12.719 | 4497 | 4.524 | 0.787 |
| 843.01 | 5881688 | 15.3 | 0.56 | 104.440 | 4.190 | 5784 | 4.396 | 1.092 |
| 845.01 | 6032497 | 15.4 | 0.35 | 110.290 | 16.330 | 5646 | 4.444 | 1.224 |
| 846.01 | 6061119 | 15.5 | 1.37 | 119.713 | 27.807 | 5612 | 4.597 | 0.846 |
| 847.01 | 6191521 | 15.2 | 0.70 | 136.898 | 80.868 | 5469 | 4.559 | 1.894 |
| 849.01 | 6276477 | 15.0 | 0.24 | 103.936 | 10.355 | 5303 | 4.475 | 0.956 |
| 850.01 | 6291653 | 15.3 | 0.89 | 109.522 | 10.526 | 5236 | 4.549 | 0.865 |
| 851.01 | 6392727 | 15.3 | 0.50 | 102.975 | 4.583 | 5570 | 4.551 | 0.892 |
| 852.01 | 6422070 | 15.3 | 0.20 | 104.904 | 3.762 | 5448 | 4.466 | 0.980 |
| 853.01 | 6428700 | 15.4 | 0.28 | 102.690 | 8.204 | 4842 | 4.472 | 0.906 |
| 855.01 | 6522242 | 15.2 | 1.21 | 128.787 | 41.408 | 5316 | 4.586 | 0.832 |
| 856.01 | 6526710 | 15.3 | 0.91 | 105.855 | 39.749 | 5858 | 4.592 | 0.861 |
| 857.01 | 6587280 | 15.1 | 0.19 | 107.884 | 5.715 | 5033 | 4.629 | 0.764 |
| 858.01 | 6599919 | 15.1 | 0.86 | 106.989 | 13.610 | 5440 | 4.450 | 0.999 |
| 863.01 | 6784235 | 15.5 | 0.22 | 105.152 | 3.168 | 5651 | 4.593 | 0.851 |
| 865.01 | 6862328 | 15.1 | 0.63 | 155.237 | 119.021 | 5560 | 4.704 | 1.232 |
| 867.01 | 6863998 | 15.2 | 0.32 | 113.274 | 16.086 | 5059 | 4.521 | 0.881 |
| 868.01 | 6867155 | 15.2 | 1.04 | 141.431 | 206.789 | 4118 | 4.517 | 0.927 |
| 871.01 | 7031517 | 15.2 | 0.91 | 112.422 | 12.941 | 5650 | 5.051 | 0.477 |
| 872.01 | 7109675 | 15.3 | 0.65 | 119.684 | 33.593 | 5127 | 4.592 | 0.810 |
| 873.01 | 7118364 | 15.0 | 0.14 | 105.226 | 4.348 | 5470 | 4.784 | 0.789 |
| 874.01 | 7134976 | 15.0 | 0.17 | 102.977 | 4.602 | 5037 | 4.561 | 0.706 |
| 875.01 | 7135852 | 15.7 | 0.34 | 103.624 | 4.221 | 4198 | 4.865 | 0.780 |
| 876.01 | 7270230 | 15.9 | 0.68 | 104.898 | 6.998 | 5417 | 4.865 | 0.589 |
| 878.01 | 7303253 | 15.3 | 0.41 | 106.808 | 23.591 | 4749 | 4.281 | 1.160 |
| 882.01 | 7377033 | 15.5 | 1.20 | 103.694 | 1.957 | 5081 | 4.572 | 0.826 |
| 883.01 | 7380537 | 15.8 | 1.05 | 103.101 | 2.689 | 4674 | 4.821 | 0.642 |
| 887.01 | 7458762 | 15.0 | 0.22 | 108.345 | 7.411 | 5601 | 4.525 | 0.923 |
| 889.01 | 757450 | 15.3 | 1.52 | 102.992 | 8.885 | 5101 | 4.480 | 0.933 |
| 890.01 | 7585481 | 15.3 | 0.84 | 109.623 | 8.099 | 5976 | 4.561 | 1.104 |
| 891.01 | 7663691 | 15.1 | 0.34 | 109.969 | 10.006 | 5851 | 4.593 | 1.244 |
| 892.01 | 7678434 | 15.2 | 0.23 | 105.617 | 10.372 | 5010 | 4.604 | 0.788 |
| 895.01 | 7767559 | 15.4 | 1.24 | 104.894 | 4.409 | 5436 | 4.372 | 1.195 |
| 900.01 | 7938496 | 15.4 | 0.45 | 105.339 | 13.810 | 5692 | 4.335 | 1.172 |
| 901.01 | 8013419 | 15.8 | 0.26 | 109.938 | 12.733 | 4213 | 4.716 | 0.359 |
| 902.01 | 8018547 | 15.8 | 0.83 | 169.808 | 83.904 | 4312 | 4.616 | 0.940 |
| 903.01 | 8039892 | 15.8 | 0.95 | 106.433 | 5.007 | 5620 | 4.776 | 1.256 |
| 906.01 | 8226994 | 15.5 | 0.23 | 107.135 | 7.157 | 5017 | 4.558 | 0.836 |
| 908.01 | 8255887 | 15.1 | 1.11 | 104.446 | 4.708 | 5391 | 4.245 | 1.288 |
| 910.01 | 8414716 | 15.7 | 0.26 | 104.720 | 5.392 | 5017 | 4.863 | 0.876 |
| 911.01 | 8490993 | 15.4 | 0.18 | 104.006 | 4.094 | 5820 | 4.783 | 0.758 |
| 912.01 | 8505670 | 15.1 | 0.22 | 104.804 | 10.849 | 4214 | 4.608 | 0.637 |
| 914.01 | 8552202 | 15.4 | 0.23 | 102.731 | 3.887 | 5479 | 4.965 | 1.126 |
| 916.01 | 8628973 | 15.1 | 0.36 | 104.312 | 3.315 | 5401 | 4.480 | 0.959 |
| 917.01 | 8655354 | 15.2 | 0.29 | 106.356 | 6.720 | 5681 | 4.478 | 0.982 |
| 918.01 | 8672910 | 15.0 | 0.99 | 139.583 | 39.648 | 5321 | 4.544 | 1.038 |
| 920.01 | 8689031 | 15.1 | 0.16 | 123.502 | 21.802 | 5330 | 4.859 | 0.608 |
| 922.01 | 8826878 | 15.4 | 0.24 | 104.624 | 5.155 | 5253 | 4.456 | 0.976 |
| 923.01 | 8883593 | 15.5 | 0.32 | 107.901 | 5.743 | 5669 | 4.596 | 1.024 |
| 924.01 | 8951215 | 15.2 | 0.36 | 106.306 | 39.478 | 5951 | 4.529 | 0.935 |
| 927.01 | 9097120 | 15.5 | 1.46 | 121.982 | 23.900 | 5957 | 4.557 | 0.903 |
| 931.01 | 9166862 | 15.3 | 1.15 | 103.679 | 3.856 | 5714 | 4.776 | 1.011 |
| 934.01 | 9334289 | 15.8 | 0.32 | 106.008 | 5.827 | 5733 | 4.655 | 0.861 |
| 935.01 | 9347899 | 15.2 | 0.40 | 113.013 | 20.859 | 6345 | 4.696 | 1.018 |
| 937.01 | 9406990 | 15.4 | 0.20 | 109.572 | 20.835 | 5349 | 4.685 | 0.725 |
| 938.01 | 9415172 | 15.6 | 0.24 | 104.701 | 9.946 | 5342 | 4.582 | 0.838 |
| 940.01 | 9479273 | 15.0 | 0.54 | 102.571 | 6.105 | 5284 | 4.629 | 1.337 |
| 942.01 | 9512687 | 15.4 | 0.23 | 107.857 | 11.515 | 4997 | 4.734 | 0.663 |
| 944.01 | 9595686 | 15.4 | 0.37 | 103.244 | 3.108 | 5166 | 4.495 | 0.921 |
| 945.01 | 9605514 | 15.1 | 0.23 | 121.860 | 25.852 | 6059 | 4.594 | 1.072 |
| 948.01 | 9761882 | 15.6 | 0.19 | 106.717 | 24.582 | 5298 | 4.946 | 0.706 |
| 949.01 | 9766437 | 15.5 | 0.27 | 103.766 | 12.533 | 5733 | 4.703 | 0.909 |
| 951.01 | 9775938 | 15.2 | 0.58 | 104.546 | 13.197 | 4767 | 4.255 | 1.205 |
| 955.01 | 9825625 | 15.1 | 0.23 | 108.731 | 7.039 | 6121 | 4.510 | 1.141 |
| 956.01 | 9875711 | 15.2 | 0.50 | 108.645 | 8.361 | 4580 | 4.334 | 1.051 |
| 152.01 | 8394721 | 13.9 | 0.57 | 91.750 | 52.088 | 6187 | 4.536 | 0.936 |
| 152.02 | 8394721 | 13.9 | 0.31 | 66.634 | 27.401 | 6187 | 4.536 | 0.936 |
| 152.03 | 8394721 | 13.9 | 0.29 | 69.622 | 13.484 | 6187 | 4.536 | 0.936 |
| 191.01 | 5972334 | 15.0 | 1.06 | 65.385 | 15.359 | 5495 | 4.519 | 0.921 |
| 191.02 | 5972334 | 15.0 | 0.19 | 65.492 | 2.419 | 5495 | 4.519 | 0.921 |
| 209.01 | 10723750 | 14.3 | 1.05 | 68.635 | 50.789 | 6221 | 4.478 | 1.418 |
| 209.02 | 10723750 | 14.3 | 0.69 | 78.821 | 18.796 | 6221 | 4.478 | 1.418 |
| 877.01 | 7287995 | 15.0 | 0.24 | 103.952 | 5.955 | 4211 | 4.566 | 0.678 |
| 877.02 | 7287995 | 15.0 | 0.21 | 114.227 | 12.038 | 4211 | 4.566 | 0.678 |
| 896.01 | 7825899 | 15.3 | 0.38 | 108.568 | 16.240 | 5206 | 4.629 | 0.821 |
| 896.02 | 7825899 | 15.3 | 0.28 | 107.051 | 6.308 | 5206 | 4.629 | 0.821 |